\documentclass[12pt,a4paper,english]{article}
\usepackage{jheppub}
\usepackage{amsfonts}
\usepackage{mathtools}
\usepackage{braket}
\usepackage{slashed}
\usepackage{multirow}
\usepackage[utf8]{inputenc}
\usepackage{amsmath}
\usepackage{esint}
\usepackage{amssymb}
\usepackage{graphicx}
\usepackage{subcaption}

\usepackage{tikz}
\usetikzlibrary{calc, positioning}
\usepackage{xcolor}

\definecolor{defaultPlotColorRG}{rgb}{0.368417,0.506778,0.709798}

\usepackage{pstricks}
\usepackage{color}
\usepackage{bbold}

\usepackage{booktabs}

\newcommand{\be}{\begin{equation}}
\newcommand{\ee}{\end{equation}}
\newcommand{\bea}{\begin{eqnarray}}
\newcommand{\eea}{\end{eqnarray}}

\newcommand{\p}{\partial}
\newcommand{\unit}{{\mathbb 1}}

\def\cR{{\cal R}}
\def\cS{{\cal S}}

\def\cZ{{\cal Z}}

\def\imath{\mathrm{i}}

\newcommand{\gb}{\bar{g}}

\newcommand{\Db}{\bar{D}}

\newcommand{\Rb}{\bar{R}}

\newcommand{\NGFP}[1]{NGFP$_{#1}$}

\usepackage{babel}

\title{On avoiding Ostrogradski instabilities within Asymptotic Safety}

\author{Daniel Becker,}
\author{Chris Ripken,}
\author{Frank Saueressig}
\affiliation{
Institute for Mathematics, Astrophysics and Particle Physics (IMAPP),\\
Radboud University Nijmegen, Heyendaalseweg 135, 6525 AJ Nijmegen, The Netherlands\\
}
\emailAdd{D.Becker@science.ru.nl}
\emailAdd{aripken@science.ru.nl}
\emailAdd{f.saueressig@science.ru.nl}

\abstract{
	We study the renormalization group flow of gravity coupled to scalar matter using functional renormalization group techniques. The novel feature is the inclusion of higher-derivative terms in the scalar propagator. Such terms give rise to Ostrogradski ghosts which signal an instability of the system and are therefore dangerous for the consistency of the theory. Since it is expected that such terms are generated dynamically by the renormalization group flow they provide a potential threat when constructing a theory of quantum gravity based on Asymptotic Safety. Our work then establishes the following picture: upon incorporating higher-derivative terms in the scalar propagator the flow of the gravity-matter system possesses a fixed point structure suitable for Asymptotic Safety. This structure includes an interacting renormalization group fixed point where the Ostrogradski ghosts acquire an infinite mass and decouple from the system. Tracing the flow towards the infrared it is found that there is a subset of complete renormalization group trajectories which lead to stable renormalized propagators. This subset is in one-to-one correspondence to the complete renormalization group trajectories obtained in computations which do not keep track of the higher-derivative terms. Thus our asymptotically safe gravity-matter systems are not haunted by Ostrogradski ghosts.
}

\keywords{quantum gravity, functional renormalization, asymptotic safety, gravity-matter systems}

\begin{document}
\maketitle


\section{Introduction}
\label{intro}
Constructing phenomenologically interesting quantum field theories which are valid at all length scales is one of the central topics in theoretical high-energy physics to date. 
For gravity, such theories may be realized through the Asymptotic Safety mechanism, see \cite{Niedermaier:2006wt,Codello:2008vh,Litim:2011cp,Reuter:2012id,Reuter:2012xf,Nagy:2012ef,Roberto:book,Bonanno:2017pkg} for reviews. 
As first suggested by Weinberg  \cite{Weinberg:1980gg,Weinproc1} this mechanism could provide a consistent and predictive quantum theory of gravity within the well-established framework of quantum field theory.
The key idea of this program is that the gravitational renormalization group (RG) flow possesses a non-trivial RG fixed point. At this fixed point (some of) the dimensionless couplings take non-zero values, so that the resulting theory is interacting. This is in contrast to the free (or Gaussian) fixed points underlying asymptotic freedom where the high-energy completion is provided by a free theory. Starting from the seminal work \cite{Reuter:1996cp}, there is, by now, substantial evidence that gravity in four spacetime dimensions actually possesses a non-Gaussian fixed point suitable for Asymptotic Safety. In particular, it has been shown that this fixed point is robust under the inclusion of the two-loop counterterm \cite{Gies:2016con} and is connected to a classical regime through a crossover \cite{Reuter:2001ag}. Besides ensuring the absence of unphysical divergences, this fixed point also comes with predictive power: any UV-repulsive direction of the fixed point allows to express the associated coupling as a function of the relevant parameters. The resulting relations may be tested experimentally, at least in principle. 

While the prospects of obtaining a quantum description of the gravitational force valid at all length scales is already intriguing, it is also clear that a realistic description of our world also requires the inclusion of matter degrees of freedom. While there has already been significant effort geared towards understanding the role of the Asymptotic Safety mechanism for gravity-matter systems, the picture is still far from complete. In order to discuss potential UV-completions of gravity-matter systems it is useful to distinguish between the two cases where the matter sector of the underlying fixed point is Gaussian or non-Gaussian in the sense that matter self-interactions are either absent or turned on. On general grounds, one may expect though that non-trivial interactions in the gravitational sector also induce non-trivial matter self-couplings, see e.g.\ \cite{Eichhorn:2012va} for a discussion. Depending on the details of the approximation used to investigate the fixed point structure of the gravity-matter system, it is conceivable that a matter fixed point which is actually non-Gaussian may be projected onto a Gaussian one if the approximation used to probe it does not include self-interactions. Conversely, a fixed point identified as Gaussian may split into a Gaussian and non-Gaussian one once additional couplings are probed.

In order to get an idea which matter sectors could actually be compatible with Asymptotic Safety, Refs.\  \cite{Dona:2013qba,Dona:2014pla,Meibohm:2015twa,Dona:2015tnf} studied projections of the full RG flow where the matter sector contained an arbitrary number of minimally coupled scalars $N_s$, vectors $N_v$, and Dirac fermions $N_D$. Complementary results for the case where spacetime carries a foliation structure have been reported in \cite{Biemans:2017zca}. While all studies agree on the statement that the matter content of the standard model of particle physics leads to a fixed point structure suitable for realizing the Asymptotic Safety mechanism, the precise values for $N_s$, $N_v$, and $N_D$ supporting a NGFP are different. Restricting to the cases where the matter sector contains scalar fields only, \cite{Dona:2013qba,Dona:2014pla,Meibohm:2015twa,Dona:2015tnf} report an upper bound $N_s \lesssim 16 - 20$, while in \cite{Biemans:2017zca} no such bound is present in agreement with the initial works \cite{Percacci:2002ie,Percacci:2003jz}. This difference can be traced back to different choices for the coarse graining operators and definitions of Newton's constant employed in these works. In particular, Refs.\ \cite{Dona:2014pla,Meibohm:2015twa,Dona:2015tnf} define Newton's constant based on the flat space graviton propagator while \cite{Dona:2013qba,Biemans:2017zca} resort to a background Newton's constant. As
argued in \cite{Meibohm:2015twa} matter degrees of freedom contribute differently in these settings. The two pictures are in qualitative agreement if $N_s$ is small  but start to deviate once the matter contribution becomes significant.

In a complementary approach, the fixed point structure arising within scalar-tensor theory has been studied in \cite{Narain:2009fy,Narain:2009gb,Benedetti:2013nya,Henz:2013oxa,Percacci:2015wwa,Labus:2015ska,Borchardt:2016pif,Henz:2016aoh}.\footnote{For related studies of RG flows of scalar field theories in a fixed (curved) background spacetime see \cite{Bridle:2013sra,Benedetti:2014gja,Shapiro:2015ova,Guilleux:2015pma}.}
This setup includes two arbitrary functions of the scalar field $\phi$, a scale-dependent scalar potential $V_k(\phi)$ and a function $F_k(\phi)$ encoding the coupling of the scalar field to the Ricci scalar. In $d=3$ this setting gives rise to a Wilson-Fisher type RG fixed point which can be understood as a gravitational-dressed version of the Wilson-Fisher fixed point known in a non-dynamical flat background. In $d=4$ the analogous analysis indentifies a fixed point with a Gaussian matter sector. In particular the scalar mass and $\phi^4$-coupling vanish at this fixed point. Ref.\ \cite{Henz:2016aoh} supplements this setting by a third scale-dependent function $K_k(\phi)$ dressing the scalar kinetic term. In this generalization also a non-Gaussian matter fixed point  has been identified.

The influence of gravity on the flow of gauge-couplings has extensively been discussed in both perturbative \cite{Robinson:2005fj,Pietrykowski:2006xy,Ebert:2007gf,Toms:2007sk} and non-perturbative \cite{Daum:2009dn,Harst:2011zx,Folkerts:2011jz,Christiansen:2017gtg,Eichhorn:2017lry} settings. Fundamental aspects related to the inclusion of fermions have been discussed in \cite{Dona:2012am,Gies:2013noa} and the compatibility of light chiral fermions with asymptotic safety has been argued in \cite{Eichhorn:2011pc,Meibohm:2016mkp,Eichhorn:2016vvy}. Starting from the prediction of the Higgs mass based on Asymptotic Safety \cite{Shaposhnikov:2009pv}, mass hierarchies in the standard model and its extensions have been studied in \cite{Eichhorn:2014qka,Eichhorn:2015kea,Eichhorn:2017ylw} while the influence of gravitational interactions on the flow of Yukawa-couplings has been studied in \cite{Zanusso:2009bs,Oda:2015sma,Eichhorn:2016esv,Hamada:2017rvn,Christiansen:2017qca,Gies:2017zwf}.\footnote{For a controlled realization of the Asymptotic Safety mechanism in gauged Yukawa-systems and their phenomenological applications see \cite{Litim:2014uca,Esbensen:2015cjw,Bond:2016dvk,Bond:2017wut}.}

Based on these works there have been several key insights related to asymptotically safe gravity-matter systems. Firstly, non-Gaussian fixed points in the matter sector may come with a higher predictive power than their Gaussian counterparts. In Ref.\ \cite{Harst:2011zx} this property has been used to predict the value of the fine-structure constants based on the Asymptotic Safety mechanism. Secondly, a non-vanishing fixed point value for the $U(1)$ hypercharge may provide a solution to the triviality problem of the standard model \cite{Christiansen:2017gtg}. Thirdly, the Higgs mass can be predicted correctly based on the beta functions of the standard model completed by the Asymptotic Safety mechanism above the Planck scale \cite{Shaposhnikov:2009pv}.

These salient features are, however, also accompanied by the lurking danger that the non-vanishing gravitational interactions may induce potentially dangerous terms in the fixed point action. Typical candidates are higher-derivative terms contributing to propagators of matter fields, which are typically associated with \mbox{Ostrogradski} \mbox{instabilities} or the violation of unitarity, see \cite{Woodard:2015zca,Crisostomi:2017aim} for reviews. In this work we initiate the study of this class of interaction terms for gravity-matter flows. For transparency we focus on the simplest possible model comprising the Einstein-Hilbert action supplemented by minimally coupled scalar fields including a higher-derivative term in the scalar propagator. We show that, as expected, the higher-derivative term is generated along the RG flow. Quite remarkably, the flow admits RG trajectories for which the ghost degrees of freedom decouple in the renormalized propagator.
These findings constitute a highly non-trivial consistency test concerning the structure of asymptotically safe gravity-matter systems. From a complementary viewpoint they also provide the initial step towards extending the classical stability analysis of Horndeski \cite{Horndeski:1974wa} and ``beyond Horndeski'' theories \cite{Langlois:2015cwa,Motohashi:2016ftl} to the quantum level. 

The remaining work is organized as follows. The Ostrogradski construction and its loop-holes are reviewed in Sect.\ \ref{sect.2}. Sect.\ \ref{sect.3} introduces the setup of our RG computation incorporating a higher-derivative kinetic term in the scalar sector, and reports the resulting beta functions in Sect.\ \ref{sect.3c}. The properties of the RG flow are investigated in Sect.\ \ref{sect.4} and we discuss the consequences of our findings in Sect.\ \ref{conclusion}. Technical details related to the evaluation of the flow equation using a non-smooth regulator are relegated to Appendix \ref{App.A}. 
\section{Higher-derivative terms and Ostrogradski instability}
\label{sect.2}
We start by briefly reviewing the classical Ostrogradski instability and its loopholes, mainly following the expositions \cite{Woodard:2015zca,Crisostomi:2017aim}.
\subsection{The instability $\ldots$}
\label{sect.2a}
It was shown by Ostrogradski in the 1850's that non-degenerate classical systems containing time derivatives of finite degree larger than two give rise to Hamiltonians whose kinetic term is not bounded from below \cite{Ostrogradsky:1850}. Irrespective of the exact form of the action, the unbounded Hamiltonian will yield several unwanted phenomena, related to the instability of the system. At the classical level, the presence of degrees of freedom coming with a wrong sign kinetic term allows to accelerate particles to infinite velocity while keeping the total energy of the system constant.

This type of instability also appears in the corresponding quantum system. While the presence of higher-derivative terms in the propagators lowers the degrees of divergencies arising in loop computations, the presence of positive and negative energy states may trigger an instantaneous decay of the vacuum. Naively, a way out may be to reinterpret the negative-energy creation and annihilation operators as positive-energy annihilation and creation operators, respectively. Although this seems to cure the instability of the vacuum state, this procedure yields states with negative norm. Removing these states from the physical spectrum, however, yields a non-unitary $S$-matrix.

In the case of a non-interacting scalar field theory, the Ostrogradski instability can be nicely illustrated by the K\"all\'en-Lehmann representation \cite{Weinberg:1995mt}. This representation expresses the dressed propagator $G(x-y)$ as a superposition of freely propagating  particles with mass $\mu \ge 0$ and propagator
\be
 G_{\text{free}}(x;\mu^2) = \int \frac{\mathrm{d}^dp}{(2\pi)^d} \, \frac{1}{p^2 + \mu^2} \, e^{ipx} \, .  
\ee
such that
\be
G(x-y)
=
\int_0^\infty	\mathrm{d}\mu^2	\,	\rho(\mu^2) \, 	G_{\text{free}}(x-y;\mu^2)
\,\text{.}
\ee
For a unitary theory, the spectral density $\rho(\mu^2)$ is a sum over norm-states with positive coefficients, thus $\rho(\mu^2)\geq 0$. If $\rho(\mu^2) < 0$ for some $\mu^2$ in the physical sector of the theory, then unitarity issues arise.

In this article, we will study a system containing scalar fields $\phi$ where the propagator contains a fourth order kinetic term (see Sect.\ \ref{sect.3a})\footnote{Throughout this work we work with Euclidean signature with a positive definite metric. Unitarity can then be studied either by performing a Wick rotation to Minkowski space, or by the Euclidean analogue, reflection positivity. The question which propagators satisfy reflection positivity is adressed in \cite{mathphyscollaboration}.}
\be\label{matterans}
S^{\rm matter} = \tfrac{1}{2}  \, Z \int \tfrac{d^dp}{(2\pi)^{d}} \, \phi \left[ p^2 + Y p^4 \right] \phi \, , 
\ee
where $Z$ denotes a wave-function renormalization and $Y$ is the coupling associated with the higher-derivative term.
This has a propagator expanded in a Fourier basis given by
\be
G(p)
=
\frac{1}{Z}	\frac{1}{p^2	+	Y	p^4} \, 
\text{.}
\ee
Using partial fraction decomposition, we can expand this in terms of free propagators:
\be\label{pfd}
G(p)
=
\frac{1}{Z}	\left(
\frac{1}{p^2}
-	\frac{1}{p^2	+	\frac{1}{Y}}
\right) \, 
\text{.}
\ee
We see that the K\"all\'en-Lehmann spectrum contains a massless state with positive density, and a state of mass 
\be\label{massrelation}
\mu^2 = Y^{-1}
\ee
with negative density. The latter state is called a (Ostrogradski) ghost. It is easy to see that the spectral density is not positive. Therefore the theory will generically be unstable.

\subsection{$\ldots$ and its loop-holes}
\label{sect.2b}
Although higher derivatives generically introduce severe fundamental flaws in a theory, there a number of ways to bypass this problem. This can be done at both  the classical and the quantum level.

One way for curing the Ostrogradski instability at the classical level is to lift the condition of non-degeneracy. In this case the higher-order time derivatives are removed by either combining them into total derivatives or using a gauge symmetry. In the former case, the  total derivatives in the Lagrangian do not contribute to the dynamics. Provided that this procedure removes all higher-derivative terms, this results in a healthy theory.\footnote{The point that a healthy theory has to remove the entire tower of higher-derivative terms has been stressed in \cite{Motohashi:2014opa}. We are greatful to H.\ Motohashi for bringing this work to our attention.} In the latter case, gauge symmetry can be used to impose an extra condition to the equations of motion. If these constraints remove the higher derivatives, the instability is cured as well.

A second option consists of replacing the terms appearing in the straight bracket of eq.\ \eqref{matterans} by an entire function of the momentum possessing a single pole of first order. This strategy results in a non-local theory which contains time-derivatives of infinite order. In this case the propagator does not admit a partial fraction decomposition and the absence of poles in the physical spectrum implies that the theory is still stable. However, the question if the resulting non-local theory is well-posed is subtle. An exposition on the treatment of this class of theories is given in \cite{Barnaby:2007ve,Barnaby:2008tc}.\footnote{For a more detailed discussion of infinite-order theories in the context of gravity we refer to \cite{Modesto:2011kw,Modesto:2014lga,Giaccari:2015vfh,Modesto:2017sdr}.}

When assessing the stability of a higher-derivative theory at the quantum level, the situation becomes even more involved. In this case the dressed propagator of the theory can be obtained from the effective action $\Gamma$ and one expects that for a stable theory this propagator does not give rise to Ostrogradski ghosts. Following the discussion of the classical case above, this may be realized in two ways:
\begin{enumerate}
	\item[a)] pushing the mass of the Ostrogradski ghost to infinity.
	\item[b)] completing the dressed propagator into an entire function.
\end{enumerate}

The first case can be illustrated by considering the action \eqref{matterans}. At the quantum level the coupling $Y$ will depend on the renormalization group scale $k$, which we indicate by $Y_k$. The requirement that the higher order derivative term does not contribute to the dressed propagator corresponds to demanding that $\lim_{k \rightarrow 0} Y_k \to 0$. At the level of the decomposition \eqref{pfd}, sending $Y \to 0$ means that the ghost mass goes to infinity. The ghost then decouples from the spectrum of the theory and does not entail an instability.\footnote{For a similar discussion in the context of higher-derivative gravity see \cite{Benedetti:2009rx}.} This scenario may be realized in two ways. Firstly, the system may exhibit a fixed point located at $Y_* = 0$. The theory at the fixed point is scale invariant and ghost-free. Secondly, an RG trajectory may be attracted to the $Y_k =0$ hyperplane as $k \rightarrow 0$. The ghost will drop out of the effective propagator rendering the renormalized theory  effectively ghost-free. 

When investigating case a), gravity plays an essential role. In its absence, the action \eqref{matterans} describes a one-parameter family of non-interacting theories parameterized by $Y$. The only ghost-free theory in this set is $Y=0$. This picture changes once a minimal coupling to the gravitational field is included. In this case the gravitational interactions induce a non-trivial flow of $Y_k$, opening the door to the nontrivial scenarios described above. 

At this stage the following remarks are in order. Firstly, we stress that the condition that the theory should be ghost-free applies to the dressed propagator (obtained at $k=0$) only. At finite values of $k$ it is expected that the process of integrating out quantum fluctuations mode-by-mode will generate higher-order derivative terms in the intermediate description. \emph{This does not signal the sickness of the theory, as its degrees of freedom should be read off from the dressed propagator.} Secondly, investigating the case b) will require generalizing the simple ansatz \eqref{matterans} to a scale-dependent function of the momentum. In \cite{Codello:2010mj} it has been shown that this class of models suffices to obtain the Polyakov effective action from a renormalization group computation. This generalization is beyond the present work though, so we will not discuss this case in detail.

\section{RG flows including higher-derivative propagators}
\label{sect.3}
Following up on the general discussion of Sect.\ \ref{sect.2}, we now perform a RG computation determining the scale-dependence of the higher-derivative coupling $Y$ in a gravity-matter setting. The key results of this section are the beta functions \eqref{betagrav}, \eqref{etaN}, \eqref{etas} and \eqref{betay} which govern the RG flow of our projection.  
\subsection{The functional renormalization group equation and its projection}
\label{sect.3a}
Currently, the predominant tool for investigating the fixed point structure and RG flows of gravity and gravity-matter systems is the functional renormalization group equation (FRGE) for the effective average action $\Gamma_k$ \cite{Wetterich:1992yh,Morris:1993qb,Reuter:1993kw,Reuter:1996cp}
\be\label{FRGE}
\p_t \Gamma_k = \frac{1}{2} {\rm Str} \left[
\left( \Gamma_k^{(2)} + \cR_k \right)^{-1} \p_t \cR_k
\right] \, .
\ee
Here $t \equiv \ln(k/k_0)$ denotes the logarithmic RG scale, $\Gamma_k^{(2)}$ is the second variation of $\Gamma_k$ with respect to the fluctuation fields and ${\rm Str}$ contains an integral over loop momenta and a sum over component fields. The regulator $\cR_k$ provides a mass-term for fluctuation modes with momenta $p^2 \lesssim k^2$ and vanishes for $p^2 \gg k^2$. The interplay of the $\cR_k$-terms in the numerator and denominator then ensures that the RG flow of $\Gamma_k$ is actually driven by quantum fluctuations with momentum scale $p^2 \approx k^2$. In this way the FRGE realizes Wilson's picture of renormalization where the RG flow is generated by integrating out fluctuations shell-by-shell in momentum space.    

The FRGE comes with some highly desirable properties. Firstly, it allows the computation of RG flows without specifying a fundamental action a priori. This feature makes the equation tailor-made for identifying interacting renormalization group fixed points. Moreover, the regulator $\cR_k$ vanishes for $k=0$ so that all quantum fluctuations are integrated out as $k\to 0$. As a consequence the effective average action agrees with the standard effective action in this limit, $\lim_{k\to 0}\Gamma_k\equiv \Gamma$. Finally, the framework turns out to be sufficiently flexible to probe settings where different classes of metric fluctuations are admitted by either implementing a linear split \cite{Reuter:1996cp}, an exponential split \cite{Ohta:2016npm,Ohta:2016jvw}, or an ADM split \cite{Manrique:2011jc,Biemans:2016rvp,Houthoff:2017oam} of the gravitational degrees of freedom. Throughout this work, we will implement a linear split, decomposing the physical metric $g_{\mu\nu}$ into a fixed background metric $\gb_{\mu\nu}$ and fluctuations $h_{\mu\nu}$ according to
\be
g_{\mu\nu} = \gb_{\mu\nu} + h_{\mu\nu} \, . 
\ee
Covariant objects carrying a bar are then constructed from the background metric while unbarred ones are constructed from $g_{\mu\nu}$. Furthermore, we will set $d=4$ throughout. While the generalization to general dimension $d$ is straightforward the rather lengthy nature of the beta functions in the general case obscures the relevant structures, so that we make this choice for clarity.

A common technique for finding non-perturbative approximate solutions of the FRGE consists of making an ansatz for $\Gamma_k$, including the operators of interest, and subsequently projecting the full flow onto the subspace spanned by the ansatz. The beta functions governing the scale-dependence of the couplings contained in the ansatz are then read off from the coefficients multiplying the interaction terms contained in the ansatz. In order to study the effects of higher-derivative terms appearing in the scalar propagators of gravity-matter systems, we make the following ansatz for the effective average action
\be\label{ans1}
\Gamma_k[g,\phi,\bar{c},c;\gb] \approx \Gamma_k^{\rm grav}[g] + \Gamma_k^{\rm matter}[\phi,g] + \Gamma_k^{\rm gf}[g;\gb] + S^{\rm ghost}[g,\bar{c},c;\gb] \, .
\ee
The gravitational part of this ansatz is taken of Einstein-Hilbert form
\be
\Gamma_k^{\rm grav}[g] = \tfrac{1}{16 \pi G_k} \int d^4x \sqrt{g} \left[
- R + 2 \Lambda_k
\right] \, .
\ee
It includes a scale-dependent Newton's constant $G_k$ and cosmological constant $\Lambda_k$. The gravitational sector is supplemented by a gauge-fixing action $\Gamma_k^{\rm gf}$ and a ghost term $S^{\rm ghost}[g,\bar{c},c;\gb]$. In order to facilitate the comparison with the results reported in \cite{Dona:2013qba}, we implement the harmonic gauge
\be
\Gamma_k^{\rm gf} = \frac{1}{32 \pi G_k} \int d^4x \sqrt{\gb} \, F_\mu \gb^{\mu\nu} F_\nu \, , \qquad \mbox{with} \qquad F_\nu = \Db^\mu h_{\mu\nu} - \tfrac{1}{2} \Db_\nu h \, .
\ee
This gauge-fixing is accompanied by a standard ghost-term\footnote{As compared to \cite{Dona:2013qba}, we do not include an anomalous dimension for the ghost fields. Thus our results correspond to $\eta_c = 0$ in \cite{Dona:2013qba}.}
\be
S^{\rm gh} = - \sqrt{2} \int d^4x \sqrt{\gb} \, \bar{c}_\mu \left[ \Db^\rho \gb^{\mu\kappa} g_{\kappa\nu} D_\rho + \Db^\rho \gb^{\mu\kappa} g_{\rho\nu} D_\kappa - \Db^\mu \gb^{\rho\sigma} g_{\rho\nu} D_\sigma \right] c^\nu \, .
\ee
The gravitational part of $\Gamma_k$ is supplemented by $N_s$ scalar fields,
\be\label{scalaraction}
\Gamma_k^{\rm matter}[\phi,g] = \tfrac{1}{2} \, Z_k \, \sum_{i=1}^{N_s} \int d^4x \sqrt{g} \, \phi^i \, \left[ \Delta + Y_k \, \Delta^2 \right] \, \phi^i \, ,
\ee
where $\Delta \equiv - g^{\mu\nu} D_\mu D_\nu$ is the Laplacian constructed from the full metric. Besides a wave-function renormalization $Z_k$, this ansatz contains a scale-dependent coupling $Y_k$ associated with a higher-derivative contribution to the scalar propagator.

\subsection{Evaluating the flow equation}
\label{sect.3c}
Starting from the ansatz \eqref{ans1}, the goal is to find the beta functions determining the scale-dependence of $G_k, \Lambda_k$ and $Y_k$ as well as the scalar anomalous dimension $\eta_s = - \p_t \ln Z_k$. This information is obtained by substituting the ansatz into the FRGE and extracting the relevant interaction terms from the trace appearing on the right-hand-side. The explicit evaluation of this operator trace requires specifying the regulator function $\cR_k$. Throughout this work, we will resort to a Litim-type profile function \cite{Litim:2000ci,Litim:2001up}, $r(z) = (1-z)\Theta(1-z)$, so that 
\be\label{regulator}
\cR_k = \cZ_k \, k^2 \, r(\Box/k^2) \, . 
\ee 
The matrix-valued wave function renormalization $\cZ_k$ is obtained from the substitution rule $\Box \mapsto P_k \equiv \Box + k^2 r(\Box/k^2)$. Following the nomenclature introduced in \cite{Codello:2008vh}, the coarse graining operator $\Box$ is chosen either as
\be\label{regulators}
\begin{split}
	{\rm Type \; I:} \qquad & \Box = \Delta \, , \\
	{\rm Type \; II:} \qquad & \Box = \Delta + q \Rb \, ,
\end{split}
\ee
where the endomorphism $E \equiv q \Rb$ is chosen such that all curvature terms appearing in $\Gamma_k^{(2)}$ become part of the coarse-graining operator. Using the Litim-profile in the regulating procedure has the advantage that all operator traces relevant in this work can be performed analytically. This comes at the price that the regulator is not smooth and the extraction of external momenta from the traces is non-trivial. In particular, contributions arising at the boundary of the momentum integrals have to be taken into account carefully. Our strategy for incorporating such terms is explained in detail in Appendix \ref{App.A}.

The projection of the operator trace is then done as follows. The flow of $G_k$ and $\Lambda_k$ can be read off from the terms proportional to $\int d^4x \sqrt{\gb} \Rb$ and $\int d^4x \sqrt{\gb}$, respectively. These contributions are conveniently found by selecting $\gb_{\mu\nu}$ as the metric on a 4-sphere and taking the background value of the scalar field $\bar{\phi} = 0$. The resulting operator traces can then be evaluated using standard heat-kernel techniques \cite{Reuter:1996cp,Codello:2008vh,Roberto:book}. In this way, one arrives at the beta functions for the gravitational couplings given in eq.\ \eqref{betagrav}.

\begin{figure}[t!]
	\centering
	\includegraphics{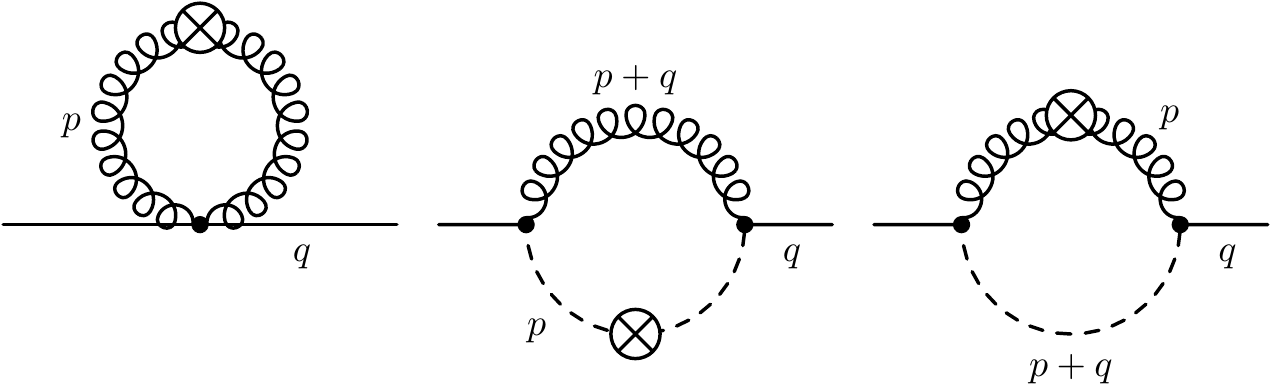}
	\caption{\label{fig:feynman} Feynman diagrams encoding the scalar contributions to the beta functions. The solid line denotes the background scalar field $\bar{\phi}$ , the dashed line denotes the fluctuating scalar field $\hat{\phi}$ , the curly line the graviton propagator. The crossed circle denotes the insertion of the cutoff operator $\p_t \mathcal{R}_k$.}
\end{figure}
The flow in the scalar sector is efficiently computed on an Euclidean background geometry $\gb_{\mu\nu} = \delta_{\mu\nu}$ and by expanding the background scalar field $\bar{\phi}(x)$ in terms of Fourier modes. Setting the fluctuation fields to zero, the scalar sector appearing on the left-hand side of the flow equation is
\be
\left. \Gamma_k^{\rm scalar} \right|_{\hat{\phi} = 0} = \tfrac{1}{2} \, Z_k \, \int \frac{d^4q}{(2\pi)^4} \, \bar{\phi}(-q^2) \left(q^2 + Y_k \, q^4 \right) \bar{\phi}(q^2) \, .  
\ee
Thus the scale-dependence of $Z_k$ and $Y_k$ is encoded in terms coming with two powers of the background scalar field and two and four powers of the momentum $q$, respectively. The Feynman diagrams generating these structures are depicted in Fig.~\ref{fig:feynman}. They consist of a pure graviton tadpole, and two diagrams with scalar-graviton loop formed by connecting two three-point vertices. The projection of the flow equation then requires extracting the contributions proportional to $q^2$ and $q^4$ from these diagrams. Following the procedure described in Appendix \ref{App.A}, this results in eqs.\ \eqref{etas} and \eqref{betay}.

The result of these computations is conveniently expressed in terms of the dimensionless couplings
\be\label{dimless}
g_k \equiv G_k \, k^2 \, , \qquad
\lambda_k \equiv \Lambda_k \, k^{-2} \, , \qquad
y_k \equiv Y_k \, k^2 \, ,
\ee
and the anomalous dimension of Newton's constant and of the scalar field
\be
\eta_N \equiv -(G_k)^{-1} \p_t G_k \, , \qquad \eta_s \equiv - (Z_k)^{-1}	\p_t Z_k \, .
\ee
The scale-dependence of the dimensionless couplings \eqref{dimless} is encoded in the beta functions which we define according to
\be\label{beta1}
\p_t g_k = \beta_g(g,\lambda,y) \, , \qquad \p_t \lambda_k = \beta_\lambda(g,\lambda,y) \, , \qquad \p_t y_k = \beta_y(g,\lambda,y) \, .
\ee
For the dimensionless variables, the system of differential equations is autonomous in the sense that the beta functions are independent of $k$.

The explicit expressions for the beta functions in the gravitational sector are
\be\label{betagrav}
\begin{split}
	\beta_g = & (2 + \eta_N) \, g \, ,  \\
	\beta_\lambda = & (\eta_N - 2) \lambda + \tfrac{g}{48 \pi} \left( \tfrac{120 }{1-2 \lambda} - \tfrac{20 \eta_N}{1-2 \lambda} - 96 + 2 N_s (6 - \eta_s) + N_s \tfrac{3 \beta_y +  y (6 - \eta_s)}{1 + y} \right) \, .
\end{split}
\ee
The anomalous dimension of Newton's constant is $y$ and $N_s$ dependent.
Inspired by \cite{Reuter:1996cp}, it can be cast into the following form:
\be\label{etaN}
\eta_N(g,\lambda,y) = \frac{g \, (B_1(\lambda) + N_s \, B_3(\lambda,y))}{1 - g B_2(\lambda) } \, .
\ee
The functions $B_1$ and $B_2$ encode the contribution of the gravitational sector. For a Type I regulator, these functions have been determined in the seminal paper \cite{Reuter:1996cp}.
For a Litim-type regulator, they read
\be\label{B1TypeI}
\begin{split}
	B_1^{\rm Type \, I} = &\,  \tfrac{1}{3\pi} \left(  \tfrac{5}{1-2\lambda} - \tfrac{9}{(1-2\lambda)^2} - 7 \right) \, , \qquad
	B_2^{\rm Type \, I} =  - \tfrac{1}{12\pi} \left(  \tfrac{5}{1-2\lambda} -  \tfrac{6}{(1-2\lambda)^2}  \right) \, .
\end{split}
\ee
For the Type II regulator, cf.\ eq.\ \eqref{regulators}, these functions become 
\be\label{B1TypeII}
\begin{split}
	B_1^{\rm Type \, II} = & \, - \tfrac{1}{3\pi} \left(  \tfrac{13}{1-2\lambda}  + 10 \right) \, , \qquad
	B_2^{\rm Type \, II} =   \tfrac{1}{12\pi} \tfrac{13}{1-2\lambda}  \, .
\end{split}
\ee
Besides the gravitational self-interaction, there is a contribution of the scalar sector to the running of $\lambda$ and $g$. 
For the latter, the additional scalar part is captured by 
\be
\begin{split}
	B_3 = & \tfrac{1}{72 \pi} \left( 12 - 3 \eta_s + \tfrac{4 \beta_y + (4-\eta_s) \, y }{1+y}\right) \, .
\end{split}
\ee
In absence of higher derivative terms in the action, i.e. $y=0$ and $\beta_y=0$ and setting the relevant ghost contributions to zero, this result agrees with \cite{Dona:2013qba}. Note that the choice of regulator, eq.\ \eqref{regulators}, enters into $B_1$ and $B_2$ only.

Next, we turn to the beta functions of the scalar sector.
The anomalous dimension for the scalar field can be expressed as
\be\label{etas}
\begin{split}
	\eta_s = \frac{g}{1-g S_4} \left( S_1 + \eta_N \, S_2 + \beta_y \, S_3 \right) \, .
\end{split}
\ee
where the $\lambda$ and $y$ dependent coefficients are given by
\be
\begin{split}
	S_1 = & \, \tfrac{1}{105 \pi} \,
	\tfrac{1}{1-2\lambda} \,
	\left( \tfrac{2}{(1+y)^2}  + \tfrac{1}{1+y} -73 - 72 \, y \right)
	- \tfrac{1}{15\pi} \tfrac{1}{(1-2\lambda)^2}
	\left( \tfrac{1}{1+y} + 9 - 4 \, y \right)
	\, , \\
	S_2 = & \, \tfrac{1}{60 \pi} \, \tfrac{1}{(1-2\lambda)^2} \,
	\left( \tfrac{1}{1+y} + 4 - 3 \, y \right) \, , \\
	S_3 = & - \tfrac{1}{7 \pi} \, \tfrac{1}{1-2\lambda} \, \left( \tfrac{1}{6(1+y)^2} + \tfrac{11}{30(1+y)} + \tfrac{2}{5}\right) \, ,  \\
	S_4 = & \, - \tfrac{1}{35 \pi} \tfrac{1}{1-2\lambda} \, \left( \tfrac{1}{4(1+y)^2} - \tfrac{1}{6(1+y)} - 3 - 2 \, y \right) \, .
\end{split}
\ee
The system is completed by the beta function for the higher-derivative coupling $y$.
Its general structure follows a similar pattern as $\eta_s$:
\be\label{betay}
\begin{split}
	\beta_y = & \frac{1}{1 - g \, S_8} \Big( \left(2+\eta_s\right) y + g \left( S_5 + \eta_N \, S_6 + \eta_s \, S_7 \right) \Big) \, .
\end{split}
\ee
The functions $S_5$ to $S_8$ depend on $\lambda$ and $y$ and are found to be
\be
\begin{split}
	S_5 = & \, \tfrac{1}{15\pi} \, \tfrac{1}{1-2\lambda} \left(  \tfrac{12}{(1+y)^2}  - \tfrac{44}{1+y} + 32\right) - \tfrac{1}{30\pi} \tfrac{1}{(1-2\lambda)^2} \left( \tfrac{35}{1+y} - 25 + 85 \, y \right) \, , \\
	S_6 = & \tfrac{1}{12\pi} \tfrac{1}{(1-2\lambda)^2}
	\left( \tfrac{3}{1+y} - 5 + 5 \, y \right) \, , \\
	S_7 = & - \tfrac{1}{30\pi} \, \tfrac{1}{1-2\lambda}
	\left(  \tfrac{3}{(1+y)^2} - \tfrac{11}{1+y} + 8 \right) \, , \\
	S_8 = & \, - \tfrac{4}{15 \pi} \tfrac{1}{1-2\lambda} \left( \tfrac{1}{(1+y)^2} -   \tfrac{1}{1+y}  \right) \, .
\end{split}
\ee
Eqs.\ \eqref{betagrav}, \eqref{etaN}, \eqref{etas} and \eqref{betay} form an implicit system which can be solved for the beta functions $\beta_\lambda$, $\beta_y$ and anomalous dimensions $\eta_N$ and $\eta_s$. In absence of the higher-derivative terms in the scalar propagator, which can be switched off by setting $y=0$ and $\beta_y = 0$, the beta functions agree with the ones reported in \cite{Dona:2013qba}. This provides a non-trivial crosscheck of our derivation.

\subsection{Structural properties of the beta functions}
The system of beta functions \eqref{betagrav}, \eqref{etaN}, \eqref{etas} and \eqref{betay} possesses several interesting properties. Firstly, $\eta_s$ and $\beta_y$ depend on the number of scalar fields $N_s$ only implicitly. This feature is readily deduced from the Feynman diagrams in Fig.\ \ref{fig:feynman} which do not contain closed scalar loops that could give rise to terms proportional to $N_s$. The number of scalars then enters the flow in the scalar sector only indirectly through the value of the cosmological constant and the anomalous dimension of Newton's constant. This suggests that the fixed point structure and flow pattern obtained from the beta functions will be rather stable under a change of the number of scalar fields.

\begin{figure}[t]
	\begin{center}
		\includegraphics{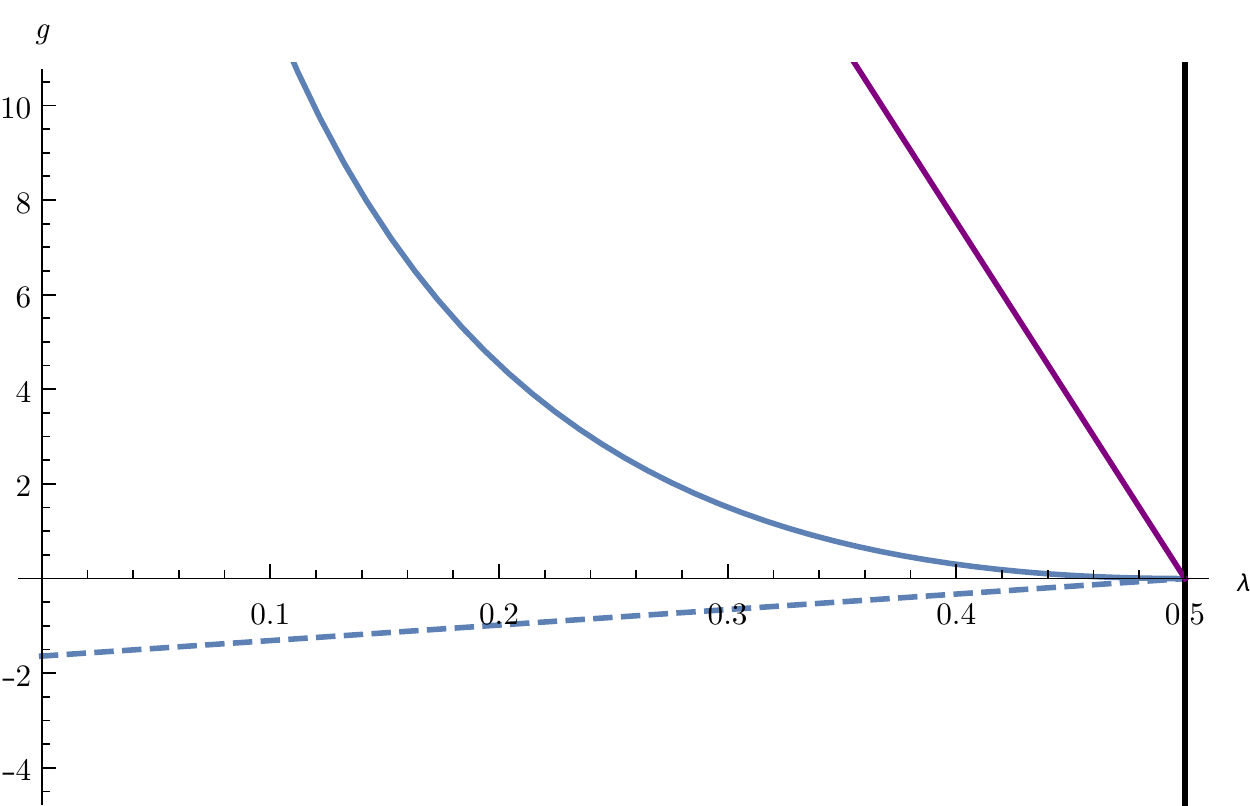}
	\end{center}
	\caption{Illustration of the singularity structure of the beta functions \eqref{betagrav}, \eqref{etaN}, \eqref{etas} and \eqref{betay} projected onto the $y=0$-plane. The black line indicates the fixed singularity at $\lambda^{\rm sing} = 1/2$. At the blue and purple lines the anomalous dimensions $\eta_N$ and $\eta_s$ diverge respectively. The solid lines apply to the Type I regulator while the dashed result is obtained from the Type II regularization procedure.
		\label{fig:sing}}
\end{figure}

Moreover, the beta functions possess several singular loci where either a beta function or an anomalous dimension diverges. The projection of these singular lines onto the $y = 0$-plane is shown in Fig.\ \ref{fig:sing}.
Inspecting $\beta_\lambda$ and $\beta_y$ one encounters two singular lines
\be
\lambda^{\rm sing} = \tfrac{1}{2} \quad \mbox{and} \qquad y^{\rm sing} = - 1 \, ,
\ee
where the denominators in the beta functions vanish.\footnote{The singularity $\lambda^{\rm sing}$ has recently been discussed in \cite{Wetterich:2017ixo}, where it has been proposed that the strong RG flows in its vicinity could drive the value of the renormalized cosmological constant to zero dynamically.} In addition one obtains singular lines when the anomalous dimensions $\eta_N$ or $\eta_s$ develop a singularity. For $\eta_N$ this locus is independent of $y$ and $N_s$ and implicitly parameterized by the relation
\be\label{etaNsing}
\eta_N^{\rm sing}: \qquad g B_2(\lambda) = 1 \, . 
\ee
Since $B_2(\lambda)$ depends on the choice of coarse-graining operator, there are two distinguished structures entailed by this relation. As illustrated in Fig.\ \ref{fig:sing} the Type I choice leads to a singular locus which screens the line $\lambda^{\rm sing} = \tfrac{1}{2}$ for positive Newton's constant while the Type II coarse graining screens $\lambda^{\rm sing} = \tfrac{1}{2}$ for $g < 0$. This observation may actually become important when ``quenching the cosmological constant'' along the lines proposed in \cite{Wetterich:2017ixo} which presupposes that an RG trajectory emanating from the classical regime can actually reach the singular locus $\lambda^{\rm sing} = \tfrac{1}{2}$.   

The hypersurface on which the scalar anomalous dimension $\eta_s$ diverges is given by a quadratic polynomial in $g$ with $\lambda$ and $y$-dependent coefficients
\be\label{etassing}
\eta_s^{\rm sing}: \qquad 1 - S_3 \, y  - g \, (S_4 + S_3 \, S_7 + S_8) + g^2 \, S_4 \, S_8  = 0 \, .
\ee
For $y=0$ the resulting line is depicted as the purple line in Fig.\ \ref{fig:sing}. The hypersurface also screens the line $\lambda^{\rm sing} = 1/2$ for $g > 0$. In the Type I coarse graining procedure $\eta_s^{\rm sing}$ is sandwiched between $\eta_N^{\rm sing}$ and $\lambda^{\rm sing} = 1/2$, while for the Type II procedure, it actually provides the screening of the $\lambda^{\rm sing} = 1/2$-line. Thus we see that the inclusion of scalar matter actually alters the singularity structure of the beta functions. At the same time, we expect that the system is rather insensitive to the inclusion of matter fields. The later point will be confirmed in more detail by the analysis of the next section.

\section{Properties of the renormalization group flow}
\label{sect.4}
We now discuss the properties of the RG flow entailed by the system \eqref{betagrav}, \eqref{etaN}, \eqref{etas} and \eqref{betay}. In Sect.\ \ref{sect.4a} we study the flow of the subsystem where the effects of the  higher-derivative terms are switched off. The results provide the basis for analyzing the effects related to the presence of higher-derivative terms in the scalar propagator in Sects.\ \ref{sect.4b} and \ref{sect.4c}. Throughout the section we focus on the flow generated by the choice \eqref{B1TypeI}, restricting ourselves to the discussion of a Type I coarse-graining operator only.

\subsection{Minimally coupled scalar fields}
\label{sect.4a}
The system \eqref{beta1} constitutes a set of autonomous coupled first order differential equations capturing the scale-dependence of $\{g_k,\lambda_k, y_k\}$.\footnote{The anomalous dimensions $\eta_N$ and $\eta_s$ can be obtained by evaluating \eqref{etaN} and \eqref{etas} along a solution of this system.} A very important concept for understanding the dynamics of such systems are its fixed points $\{g_{i,*}\}$ where, by definition,
\be\label{fpdef}
\beta_{g_i}(\{g_{j,*}\}) = 0 \, . 
\ee
The flow of the system in the vicinity of such a fixed point can be studied by linearizing the beta functions at $\{g_{i,*}\}$. The stability coefficients $\theta_i$, defined as minus the eigenvalues of the stability matrix ${\bf B}_{ij} = \left. \frac{\partial \beta_{g_j}}{\partial g_i} \right|_{g_{j,*}}$, indicate whether flows along the corresponding eigendirection are attracted  $({\rm Re}(\theta_i) > 0)$ or repelled $({\rm Re}(\theta_i) < 0)$ by the fixed point as $k \rightarrow \infty$. Eigendirections with ${\rm Re}(\theta_i) > 0$ span the UV-critical hypersurface $\cS_{\rm UV}$ of the fixed point. By construction any RG trajectory for which the fixed point provides the high-energy completion must be contained in $\cS_{\rm UV}$. The parameters pinpointing a particular RG trajectory within $\cS_{\rm UV}$ then constitute free parameters which need to be fixed by experimental data or other theoretical considerations.

\begin{figure}[t!]
	\centering
	\includegraphics[width=0.48\textwidth]{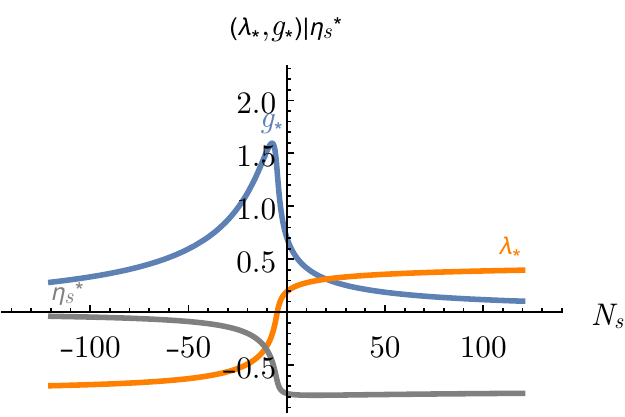}
	\includegraphics[width=0.48\textwidth]{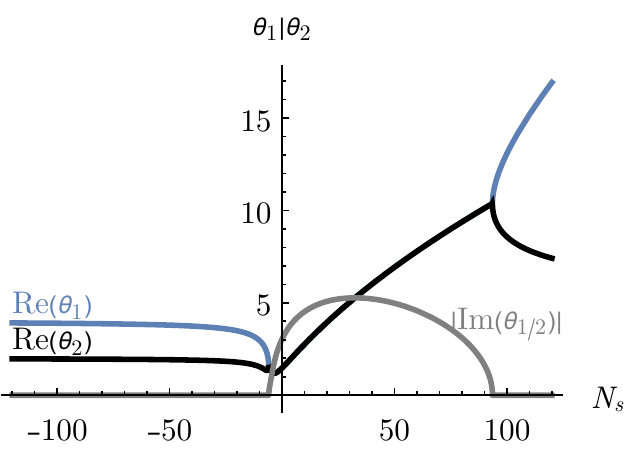}
	\caption{Characteristics of the NGFP in the minimally coupled gravity-scalar system as a function of $N_s$. 	Its position in the $\lambda$-$g$--plane and the resulting scalar anomalous dimension $\eta_s^*$ are shown in the left panel while the stability coefficients are displayed in the right panel.}
	\label{fig:kgfplocation}
\end{figure}
Before delving into the analysis of the full system, it is useful to first analyze the subsystem obtained from setting $y_k = 0$, $\beta_y = 0$. In this approximation the contributions of the higher-derivative terms in the scalar sector are switched off and the projection of the flow equation is given by the Einstein-Hilbert action supplemented by an arbitrary number $N_s$ of minimally coupled scalar fields. The RG flow resulting from similar projections has been studied in \cite{Percacci:2003jz,Dona:2013qba,Meibohm:2015twa,Dona:2015tnf,Biemans:2017zca}. The analysis of this subsection then facilitates the comparison with these works.

\begin{figure}[t!]
	\centering
	\includegraphics[width=0.48\textwidth]{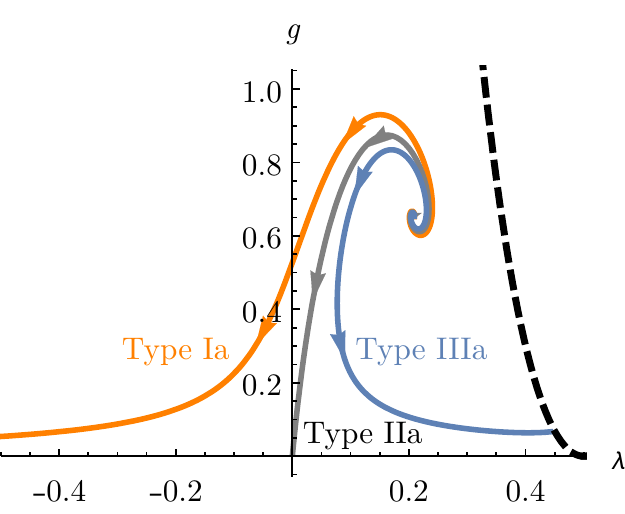}
	\includegraphics[width=0.48\textwidth]{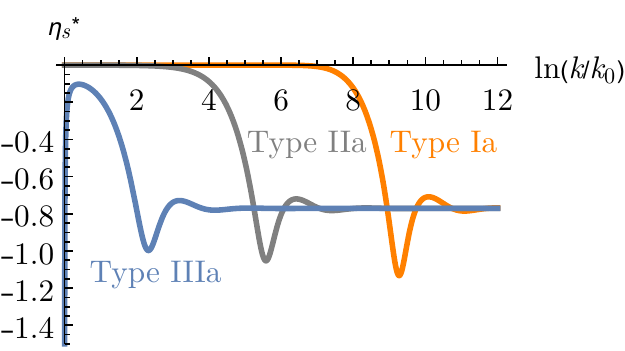}
	\caption{Three prototypical RG trajectories obtained from numerically integrating the reduced system of beta functions for $N_s = 1$ (left). The flow is governed by the interplay of the NGFP and GFP. The scalar anomalous dimension $\eta_s$ along the trajectories is shown in the right diagram. The initial scales $k_0$ are tuned such that the trajectories in the right diagram are disentangled. 
		Notice that $\eta_s$ is negative semi-definite along the entire RG flow. 
		In the UV ($k\to\infty$) the anomalous dimension $\eta_s$ approaches its fixed point value $\eta_s^* = -0.771$ independently of the specific initial conditions. In the IR $\eta_s$ remains negative and vanishes asymptotically for the solutions of Type Ia and Type IIa. Trajectories of Type IIIa terminate in the singular line $\eta_N^{\rm sing}$ triggering the divergence of $\eta_s$ at a finite value of $k$.}
	\label{fig:kganomdim}
\end{figure}

\paragraph{Fixed point structure.} The reduced system possesses two fixed points, a Gaussian and a non-Gaussian one. The Gaussian fixed point (GFP) is situated in the origin and its stability coefficients are determined by the mass-dimension of the coupling constants,
\be\label{GFP1}
(\lambda_*, g_*) =(0,0) \, , \qquad  \quad \theta_1 = 2 \, , \; \theta_2 = -2 \, . 
\ee 
The anomalous dimensions vanish at this fixed point. The stability coefficients indicate that the GFP is a saddle point in the $\lambda$-$g$--plane exhibiting one UV-attractive and one UV-repulsive eigendirection, also see the left diagram of Fig.\ \ref{fig:kganomdim}. The GFP exists for all values $N_s$.

In addition the system possesses a one-parameter family of non-Gaussian fixed points (NGFPs) parameterized by the number of scalar fields $N_s$. Its position and stability of these fixed points as a function of $N_s$ is shown in the left and right diagram of Fig.\ \ref{fig:kgfplocation}, respectively. In addition, explicit values of the position $(\lambda_*, g_*)$, the universal product $\lambda_* g_*$,  the scalar anomalous dimension evaluated at the fixed point $\eta_s^*$, and the stability coefficients for selected values of $N_s$ are provided in Table \ref{tab:kgngfp}.
Notably, there is a NGFP for all values $N_s$.  By virtue of eq.\ \eqref{betagrav} all NGFPs come with $\eta_N^* = -2$. The one-parameter family of NGFP solutions exhibits a maximal value of $g_*=1.60$ at $N_s = -7.47$. The cosmological constant $\lambda_*$ has an inflection point at $(N_s,\lambda_*)= (-5.23,-0.0399)$ and has a zero at $N_s = -4.81$. The anomalous dimension has inflection points at $(N_s,\eta_{s}^*)= (-5.46,-0.600)$ and $(N_s,\eta_{s}^*)=(35.5,-0.780)$; it has a minimum at $(N_s,\eta_{s}^*)=(14.3,-0.784)$. The \mbox{analysis} of the stability \mbox{coefficients} displayed in the right diagram of Fig.\ \ref{fig:kgfplocation} shows that all NGFPs are UV-attractive in the $\lambda$-$g$--plane. The critical exponents $\theta_i$ have a non-zero imaginary part for $N_s \in [-6 ,93]$ only. For other values of $N_s$ the critical exponents turn out to be real.
\begin{table}
	\centering
	\begin{tabular}{@{}rrrrrrrcc|c@{}}
		\multicolumn{2}{r@{}}{} \\
		\toprule
		$N_s$	 &\hspace{0.5cm} &	$g_*$		&	$\lambda_*$	&\hspace{0.5cm} & $g_* \lambda_*$ & 	$\eta_s^*$	&\hspace{0.5cm}	&	$\theta_{1}$	&	$\theta_2$	
		\\
		\midrule
		$-100$	&&	$0.333$	&	$-0.684$	&& $-0.228$ & $-0.046$	&			&	$3.898$	&	$1.963$
		\\
		$-6$	&&	$1.530$	&	$-0.111$	&& $-0.170$ &	$-0.556$	&				&	$1.822$	&	$1.475$
		\\
		$0$		&&	$0.707$	&	$0.193$	&& $0.137$ &	$-0.766$	&				&	\multicolumn{2}{r@{}}{	$1.475 \pm 3.043	\imath$	}
		\\
		$1$		&&	$0.655$	&	$0.208$ && $0.136$	&	$-0.771$	&				&	\multicolumn{2}{r@{}}{	$1.599 \pm 3.282 \imath$	}
		\\
		$10$	&&	$0.419$	&	$0.278$	&& $0.117$ &	$-0.784$	&				&	\multicolumn{2}{r@{}}{	$2.762 \pm 4.523 \imath$	}
		\\
		$100$	&&	$0.119$	&	$0.389$	&& $0.046$ &	$-0.768$	&				&	$13.09$	&	$8.572$
		\\
		\bottomrule
	\end{tabular}
	\caption{Characteristic quantities for the NGFP appearing at selected values of $N_s$. The case $N_s = 0$ corresponds to the Einstein-Hilbert truncation.}
	\label{tab:kgngfp}
\end{table}

In the interval $N_s \in [-4,16]$ the NGFP discussed above is the only non-trivial fixed point solution. Outside this window the simplified system possesses additional NGFPs. These are, however, located outside the physically interesting region located at $g > 0$ and to the left of the singular lines depicted in Fig.\ \ref{fig:sing}. Therefore, these fixed points will not be discussed in detail.

\paragraph{Flows away from the NGFP.} Beyond the vicinity of the NGFP, where the linearized approximation of the flow is valid, the RG trajectories can be constructed by integrating the beta functions of the reduced system numerically. In the case where the critical exponents of the NGFP are complex ($N_s \in [-6,93]$) the resulting phase diagram follows the same classification as in the case of pure gravity \cite{Reuter:2001ag}. For the case $N_s = 1$ three prototypical RG trajectories are shown in the left diagram of Fig.~\ref{fig:kganomdim}. The trajectories undergo a crossover from the NGFP, controlling the high-energy regime, to the GFP, controlling the classical regime of the theory.
The RG trajectory connecting the two fixed points is called ``Type IIa'' and leads to a vanishing value of the renormalized cosmological constant $\lim_{k \rightarrow 0} \Lambda_k = 0$. Trajectories flowing to the left (right) to this line are called Type Ia (Type IIIa) and give rise to a negative (positive) value of the cosmological constant in the classical regime. The present set of flow equations do not allow to continue the Type IIIa solutions to $k=0$: they terminate in the line $\eta_N^{\rm sing}$ shown in Fig.\ \ref{fig:sing} at a finite value of $k$.

The scalar anomalous dimension obtained along these sample RG trajectories is shown in the right panel of Fig.\ \ref{fig:kganomdim}. Notably $\eta_s(k) \leq 0$ along the entire flow: at the NGFP one has $\eta_s^* = -0.771$ and the scalar anomalous dimension approaches zero when the flow enters the classical regime governed by the GFP. Thus the anomalous dimension induced by the gravitational quantum corrections suppress the propagation of scalar modes on all scales. The rapid increase of $|\eta_s|$ for the Type IIIa trajectory close to its termination point is a clear indication that the present approximation is insufficient in this regime and should thus not be given too much significance. 
\subsection{Fixed point structure including higher-derivative terms}
\label{sect.4b}
We now focus on the fixed point structure of the full system \eqref{beta1} including the higher-derivative coupling $y_k$. Following the structure of the last subsection, we first discuss the fixed point structure of the system.

Inspecting the beta functions, one finds that the GFP \eqref{GFP1} has the following extension
\be\label{GFP2}
(\lambda_*, g_*,y_*) =(0,0,0) \, , \qquad  \quad \theta_1 = 2 \, , \; \theta_2 = -2 \, , \; \theta_3 = -2 \, . 
\ee 
Again there is a GFP for all values of $N_s$ and the anomalous dimensions vanish at this fixed point. The stability coefficients indicate that the GFP is a saddle point in the $\lambda$-$g$-$y$-plane exhibiting one UV-attractive and two UV-repulsive eigendirections. In particular, it may serve as an IR attractor for RG flows starting at $g_k > 0$ which subsequently leave the GFP regime along the unstable direction.

\begin{figure}[t!]
	\begin{center}
		\includegraphics[width=0.48\textwidth]{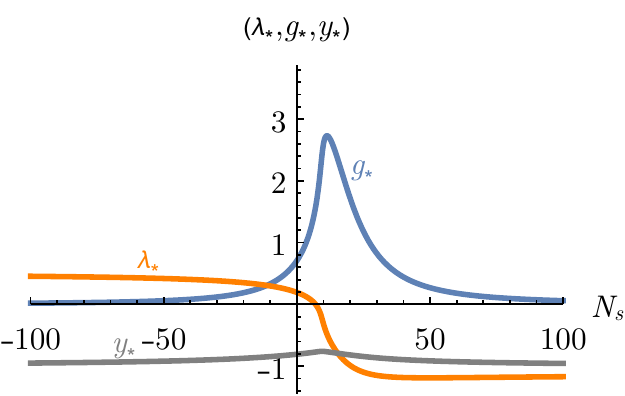}
		\includegraphics[width=0.48\textwidth]{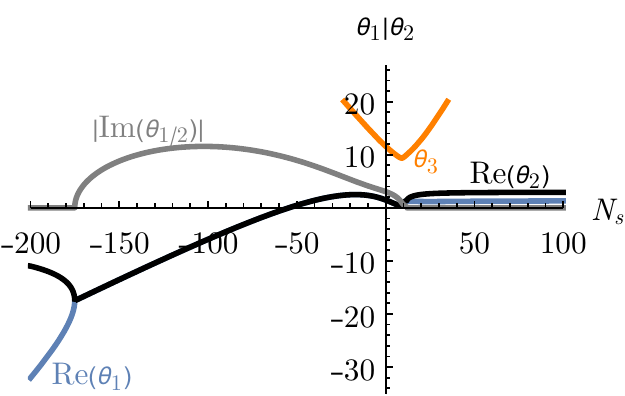} \\[3ex]
		\includegraphics[width=0.48\textwidth]{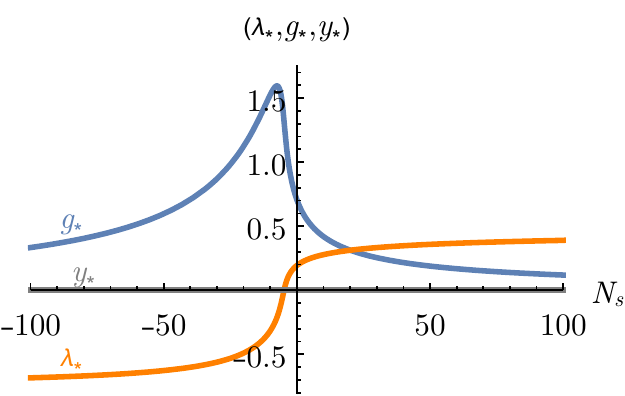}
		\includegraphics[width=0.48\textwidth]{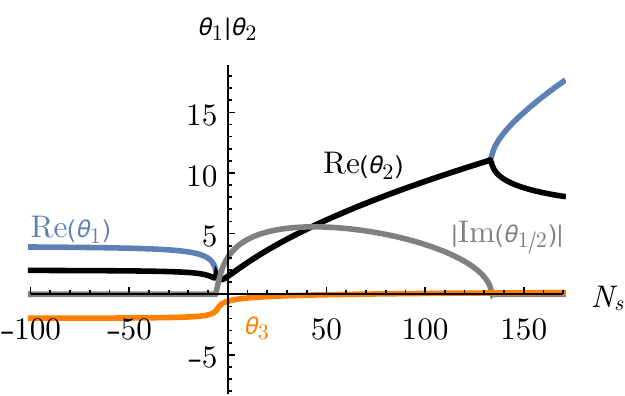} \\[3ex]
		\includegraphics[width=0.48\textwidth]{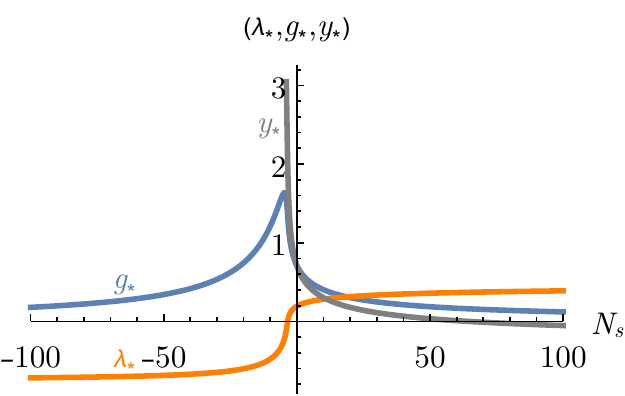}
		\includegraphics[width=0.48\textwidth]{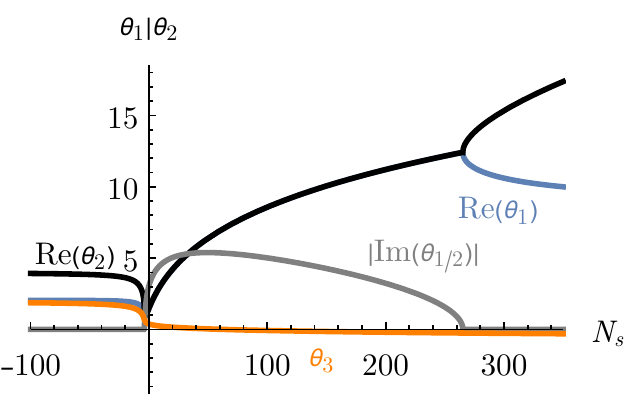}
	\end{center}
	\caption{\label{fig:NGFPvalues} Illustration of the fixed point structure resulting from the full system of beta functions \eqref{beta1} as a function of $N_s$. The characteristics of the \NGFP{-}, \NGFP{0}, and \NGFP{+} are shown in the first, second, and third row, respectively.}
	\vskip-1em
\end{figure}
The analysis of possible NGFPs starts with the following, intriguing observation: when restricted to $y = 0$, the beta function $\beta_y$, given in eq.\ \eqref{betay}, simplifies to
\be\label{betay0}
\begin{split}
	\left.	\beta_y \right|_{y=0} = & \, -	\frac{g}{6\pi} \, \frac{2+\eta_N}{(1-2\lambda)^2}
\end{split}
\ee
Thus $\beta_y$ supports a fixed point at $y_* = 0$ if $\eta_N^* = -2$. From $\beta_g$ one finds that the latter condition is precisely the anomalous dimension of Newton's coupling at any NGFP. This shows that there is an extension of the NGFP discussed in the previous section to the full system, i.e., for all values of $N_s$ we obtain a NGFP with $y_* = 0$. This family of NGFPs will be called \NGFP{0} in the sequel. Remarkably, the balancing between the anomalous dimension $\eta_N^*$ and the other contributions to $\beta_y$ works for $d=4$ only. In any other spacetime dimension the fixed point is shifted away from the $y=0$-plane.

A numerical investigation of the fixed point structure for $N_s\in [-200,350]$ \mbox{reveals} the existence of 3 families of NGFPs, parameterized by $N_s$, and located in the physically interesting region. The three families are conveniently labeled by the sign of the fixed point value $y_*$ which is either negative (\NGFP{-} branch), zero (\NGFP{0} branch), or positive (\NGFP{+} branch). The positions and stability coefficients of these fixed points are shown in Fig.\ \ref{fig:NGFPvalues}. In addition the characteristics for the NGFPs found for $N_s = 1$ are collected in Table \ref{tab:ns1fpdata}. The detailed properties of the fixed point solutions are the following.
\begin{table}[t!]
	\renewcommand{\arraystretch}{1.1}
	\begin{tabular}{@{}l|rrrrrrlc|cr@{}}
		\toprule
		&	$g_*$	&	$\lambda_*$	&	$y_*$	& \hspace{3mm}	& $g_* \lambda_*$ &	$\eta_s^*$	&	\hspace{3mm}	&	$\theta_1	$	&	$\theta_2$						&	$\theta_3$
		\\
		\midrule
		GFP			&	$0$		&	$0$			&	$0$	& & $0$		&	$0$			&					&	$+2$			&	$-2$							&	$-2$
		\\ 
		\NGFP{-}	&	$0.776$	&	$0.176$		&	$-0.804$	& & $0.137$ &	$-0.721$	&					&	\multicolumn{2}{c@{}}{	$1.34 \pm 2.92\imath$	}	&	$11.3$
		\\
		\NGFP{0}	&	$0.655$	&	$0.208$		&	$0$	& & $0.136$ 		&	$-0.771$	&					&	\multicolumn{2}{c@{}}{	$1.59 \pm 3.28\imath$	}	&	$-0.529$
		\\
		\NGFP{+}	&	$0.646$	&	$0.211$		&	$0.621$	&  & $0.136$ 	&	$-0.775$	&					&	\multicolumn{2}{c@{}}{	$1.67 \pm 3.32\imath$	}	&	$0.357$
		\\
		\bottomrule
	\end{tabular}
	\caption{Characteristic features of the four fixed points arising from the full set of beta functions \eqref{beta1} for $N_s = 1$.}
	\label{tab:ns1fpdata}
\end{table}

\paragraph{\NGFP{-}.} The characteristic properties of this family of fixed points is shown in the first line of Fig.\ \ref{fig:NGFPvalues}. Their defining criterion is that they are located at $y_* < 0$ for all values of $N_s$. More precisely,  the position $y_*$ is found to be in the interval $-1<y_*\lesssim -0.76$ and approaches the singularity $y^{\rm sing} = -1$ in the scalar propagator if $|N_s|$ becomes large. The profile for $g_*$ is peaked at $N_s \approx 11.3$ where $g_* \approx 2.73$. The cosmological constant $\lambda_*$ undergoes a crossover from $\lambda_* < 0$ for $N_s \gtrsim 7$ to $\lambda_* > 0$ for negative values $N_s$. For large negative values $N_s$ the fixed points are pushed into the corner of singular lines $\lambda^{\rm sing} = 1/2$, $y^{\rm sing} = -1$.

\begin{figure}[t!]
	\centering
	\includegraphics{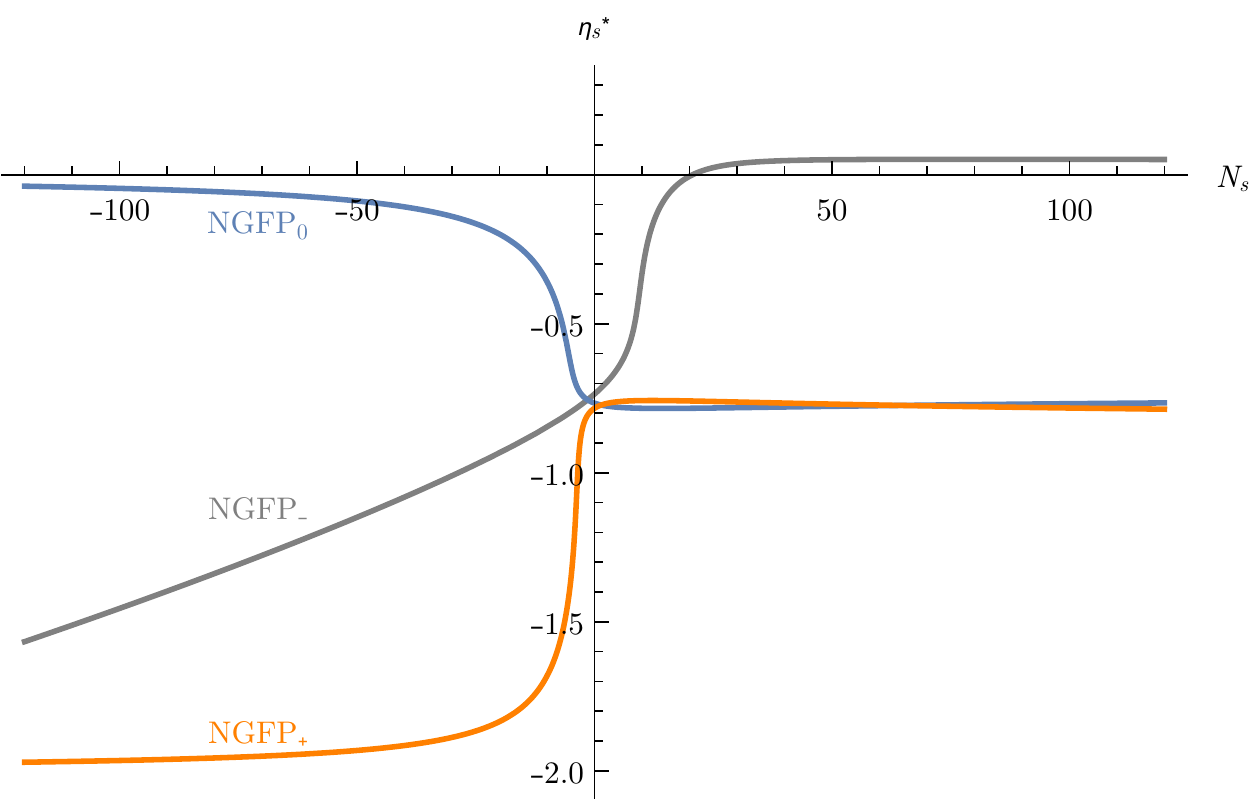}
	\caption{Fixed point value of the scalar anomalous dimension $\eta_s^*$ evaluated for the three classes of fixed points \NGFP{-}, \NGFP{0}, and \NGFP{+} as a function of $N_s$.}
	\label{fig:NGFPetaS}
\end{figure}

The stability coefficients are displayed in the upper right diagram of Fig.\ \ref{fig:NGFPvalues}. In the interval $N_s \in [-54,350]$ all three stability coefficients come with a positive real part indicating that all three couplings are 
UV-relevant. Within the interval $-175 \lesssim N_s \lesssim 12$ the two critical exponents $\theta_1$ and $\theta_2$ form a complex pair, indicating a spiraling behavior of the RG flow around \NGFP{-} in their respective directions.
Outside this window all $\theta_j$ are real valued. The scalar anomalous dimension $\eta_s^*$ is shown in Fig.\ \ref{fig:NGFPetaS}. For $N_s \lesssim 20$, we find $\eta_s^* < 0$, indicating a suppression of the scalar propagator at high energies. At $N_s \approx 20$ there is a transition to very small and positive values $\eta_s^* \lesssim 0.1$. Notably this is the only fixed point configuration where $\eta_s^*$ is actually positive.

\paragraph{\NGFP{0}.} The characteristic features of this class of fixed points is displayed in the middle line of Fig.\ \ref{fig:NGFPvalues}. All fixed points in this family are located at $y_* = 0$. Therefore this family constitutes the natural extension of the NGFP seen in the last subsection. The profiles specifying the position of these fixed points in the $\lambda_*$-$g_*$--plane resembles the one of \NGFP{-} discussed above, with the difference that their values are scaled and mirrored around $N_s\approx 0$.
This implies that the fixed point is pushed towards the singularity at $\lambda=1\slash2$ for large \emph{positive} $N_s$. 
The transition to $\lambda_*<0$ happens at negative $N_s\approx -4.81$ and the maximum value of $g_{*} \approx 1.60$ is obtained at $N_s\approx -7.47$. 

The stability properties of the fixed points \NGFP{0} can again be read off from the stability coefficients displayed in Fig.\ \ref{fig:NGFPvalues}. Two of their stability coefficients always come with a positive real part (indicating that the directions are UV-attractive). On the interval $N_s\in(-6,133]$ they form a complex conjugate pair while outside  this range both of them are real valued. The third coefficient $\theta_3$ changes sign at $N_s= 65$. For smaller values $\theta_3 < 0$, indicating that the corresponding \NGFP{0} is actually a saddle point in the $\lambda$-$g$-$y$-plane. For $N_s > 65$ all three stability coefficients have positive real parts so that the fixed points are UV-attractors in this case. The scalar anomalous dimension $\eta_s^*$ remains negative throughout and is bounded by $|\eta_s^*| < 0.77$.

\paragraph{\NGFP{+}.} The characteristic features of this class of fixed points is displayed in the bottom line of Fig.\ \ref{fig:NGFPvalues}. This class  comes with a positive $y_*$ which grows very rapidly for negative values of $N_s$. The position of the fixed points in the $\lambda$-$g$-plane are qualitatively the same as the ones found for \NGFP{0}. For large positive values $N_s\gtrsim 65$ the location $y_*$ changes sign.

In the interval $[-200,65]$, all stability coefficients appearing in this family possess a positive real part, so that the \NGFP{+} are UV-attractors in the $\lambda$-$g$-$y$--plane. Similarly to the other fixed points, the stability coefficients $\theta_1$ and $\theta_2$ form a complex pair for $-4 \lesssim N_s \lesssim 265$ and are real outside this interval . The scalar anomalous dimension $\eta_s^*$ is negative throughout and takes values between $-2 \lesssim \eta_s^* \lesssim -0.79$.

At this point the following remark is in order. Combining eqs.\ \eqref{massrelation} and \eqref{dimless}, the mass of the Ostrogradski ghost is
\begin{equation}\label{Oghostmass}
\mu^2 = \frac{k^2}{y_*} \, . 
\end{equation} 
Thus $\mu^2$ will become infinite for any RG trajectory approaching a NGFP as $k \rightarrow \infty$. This is just a consequence of the fact that a fixed point can not support a dimensionful scale. The relation \eqref{Oghostmass} also reveals that the fixed points \NGFP{0} are very special. Owed to their position at $y_* = 0$ the mass of the Ostrogradski ghost \emph{is infinite for all values $k$}. In this way, the \NGFP{0} realize the first class of loopholes discussed in sect.\ \ref{sect.2b}. Thus the extra degree of freedom is not present and one expects that the resulting theory \emph{does not suffer from an Ostrogradski instability} albeit living in a theory space which permits the presence of higher-derivative terms in the propagator a priori. 

\subsection{Phase diagram including higher-derivative terms}
\label{sect.4d}
We now extend the local analysis of the RG flow, based on its fixed point structure and stability coefficients, to a global picture. For concreteness, we focus on the case $N_s = 1$. The details of the fixed point structure arising in this setting is summarized in Table \ref{tab:ns1fpdata}. Since the essential features of the flow are set by its fixed point structure, it is clear that the analysis applies to an entire window $-6 \lesssim N_s \lesssim 12$ where the fixed point structure and stability coefficients exhibit the same qualitative behavior.

\begin{figure}[t!]
	\centering
	\includegraphics[width=0.5\textwidth]{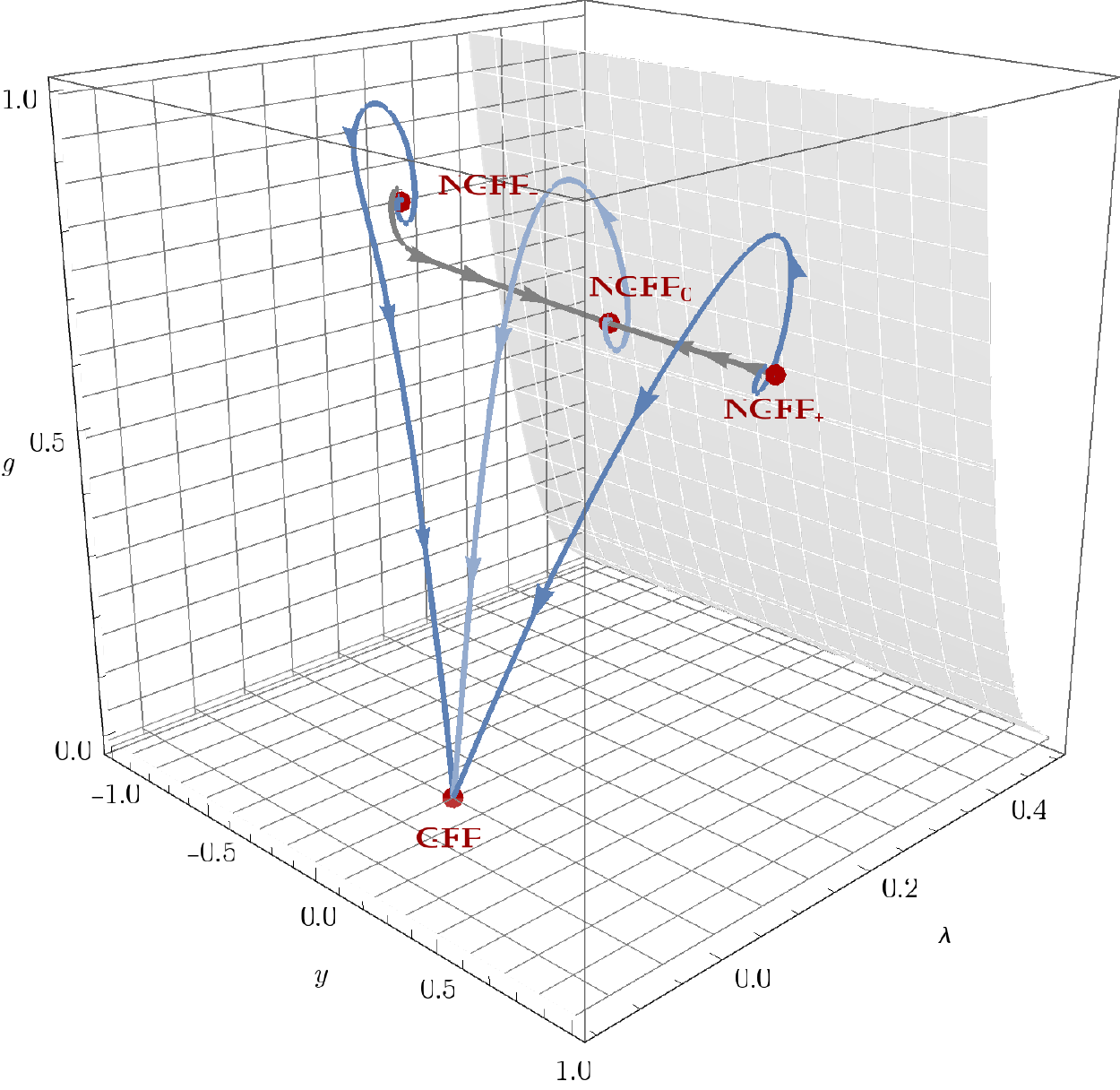}
	\caption{Illustration of the phase diagram resulting from the beta functions \eqref{beta1} for $N_s=1$. The GFP and the three NGFPs are marked with red points while the singular loci $y^{\rm sing} = -1$ and $\eta_N^{\rm sing}$ are  shaded in gray. The RG trajectories connecting the fixed points are shown in blue (Type IIa trajectories) and gray. All arrows point from UV to IR.}
	\label{fig:phasediagram}
\end{figure}
\begin{figure}[t!]
	\centering
	\includegraphics[width=0.5\textwidth]{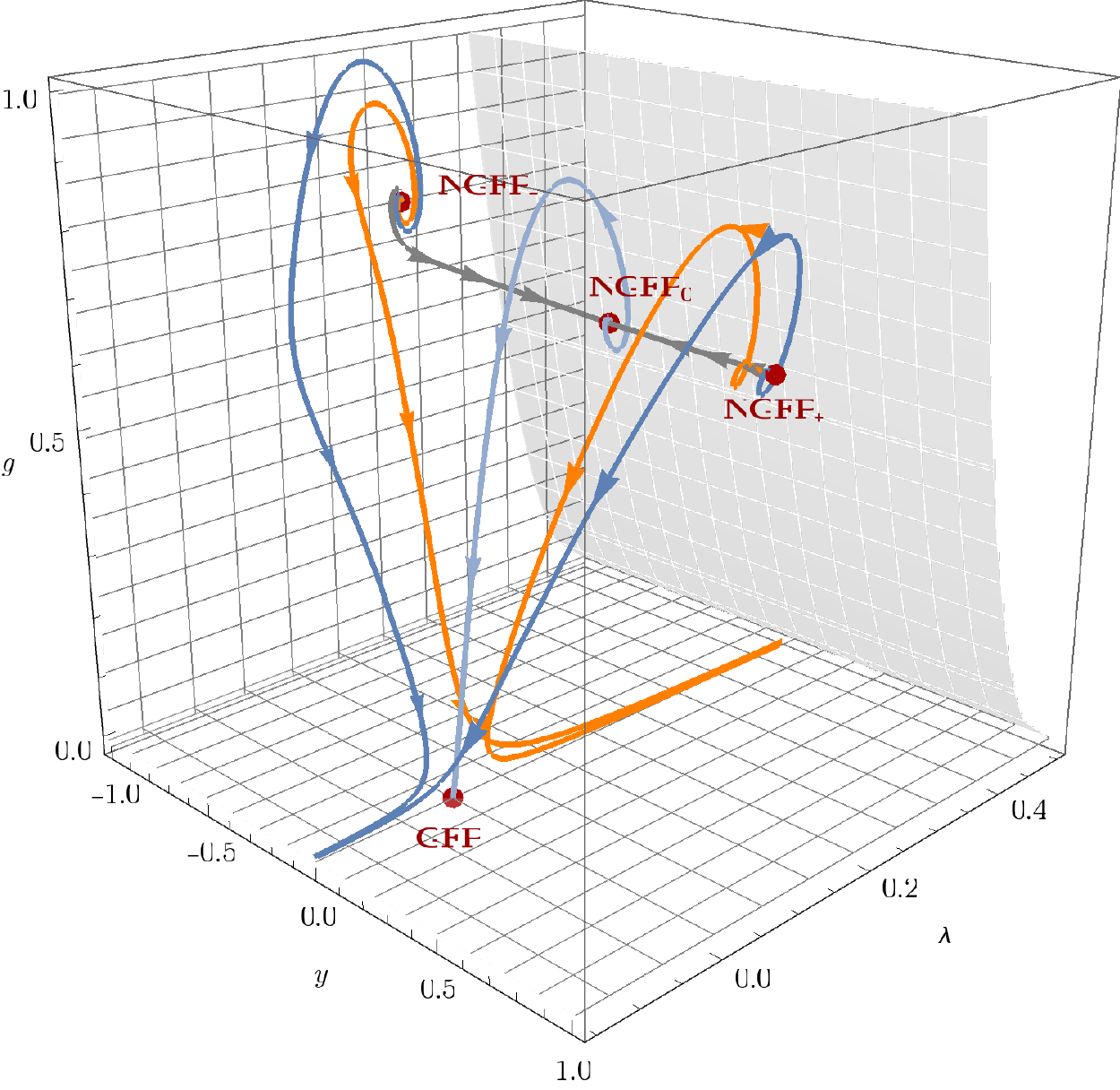}
	\caption{Illustration of the phase diagram resulting from the beta functions \eqref{beta1} for $N_s=1$. Depicted are typical RG trajectories undergoing a crossover from the NGFP to the classical regime controlled by the GFP. Depending on whether the classical value of the cosmological constant found along the flow is positive (orange curves) or negative (blue curves) the trajectories are termed Type IIIa and Type Ia, respectively. The orange solutions terminate at $\eta_N^{\rm sing}$ displayed as the gray shaded surface. All arrows point from UV to IR.}
	\label{fig:phasediagram2}
\end{figure}

The global structure of the RG flow is obtained by integrating the beta functions \eqref{beta1} numerically. A characteristic set of trajectories obtained this way is shown in Figs.\ \ref{fig:phasediagram} and \ref{fig:phasediagram2}. Fig.\ \ref{fig:phasediagram} then shows the RG trajectories connecting the 3 NGFPs (gray lines) and the NGFPs with the GFP (blue lines). Since both \NGFP{\pm} act as UV-attractors in the $\lambda$-$y$-$g$--plane and the \NGFP{0} possesses one IR-attractive eigendirection there is a single RG trajectory emanating from either \NGFP{\pm} for $k \rightarrow \infty$ and ending at the \NGFP{0} as $k\rightarrow 0$. The GFP possesses 2 IR-attractive eigendirections. As a result, one finds a unique trajectories which starts from \NGFP{0} and connects to the GFP $k\rightarrow 0$ (light blue line). This trajectory is the intersection of the two-dimensional UV-critical hypersurface of \NGFP{0} with the two-dimensional IR-critical hypersurface of the GFP. In addition there are two families of solutions which originate from \NGFP{\pm} and end at the GFP, again coming from the intersection of the 3-dimensional UV-critical hypersurfaces of the NGFPs with the IR-critical hypersurface of the GFP. These flows are exemplified by the dark blue lines. All together this set constitutes the generalization of the Type IIa trajectory displayed in Fig.\ \ref{fig:kganomdim}. 

Fig.\ \ref{fig:phasediagram2} then illustrates the generalization of the trajectories of Type Ia and Type IIIa to the $\lambda$-$y$-$g$--plane. These trajectories may emanate from all three NGFPs and subsequently cross over to the GFP. From the vicinity of the GFP they either flow to large negative values $\lambda_k$ (Type Ia) or positive $\lambda_k$ (Type IIIa) such that their projection to the $\lambda$-$g$-plane resembles the left diagram of Fig.\ \ref{fig:kganomdim}. The latter class again terminates in the hypersurface $\eta_N^{\rm sing}$ at a finite value $k$. Notably, for all physically interesting trajectories which exhibit a crossover to the GFP, $y_k$ flows to zero in the IR, provided that the underlying trajectories do not terminate at a finite value $k$. When evaluating the scalar anomalous dimension $\eta_s$ along the RG trajectories shown in Figs.\ \ref{fig:phasediagram} and \ref{fig:phasediagram2} one again obtains the qualitative behavior shown in the right diagram of Fig.\ \ref{fig:kganomdim}: for large values of $k$ $\eta_s$ is determined by its fixed point value $\eta_s^*$. Once the RG trajectory enters the vicinity of the GFP quantum effects become small, $\eta_s \ll 1$ asymptotically.

\subsection{Ghost-free RG flows in the infrared}
\label{sect.4c}
In order to determine the stability of the theory in the presence of higher-derivative terms one has  to study the renormalized scalar propagator obtained from the effective average action $\Gamma_k$ in the limit $k \rightarrow 0$. Defining $Y_0 \equiv \lim_{k \rightarrow 0} Y_k$ the (squared) mass of the Ostrogradski ghost is (cf.\ eq.\ \eqref{massrelation})
\be\label{renghostmass} 
\mu^2 =  \frac{1}{Y_0}
\ee
Hence instability will disappear from the spectrum if $Y_0 = 0$. Thus the focus of the investigation is on the IR behavior of $y_k$. Fig.\ \ref{fig:phasediagram} demonstrates that all physically interesting RG trajectories have the property that the dimensionless coupling $y_k$ goes to zero in the IR. This leaves three potential scenarios for the dimensionful coupling $Y_k = y_k k^{-2}$:  
\begin{enumerate}
	\item The dimensionless coupling $y_k$ approaches zero slower than quadratically. The canonical scaling of $Y_k$ will dominate the flow and $Y_0$ diverges. In this case the ghost becomes massless and eats up the scalar degree of freedom, see eq.\ \eqref{pfd}.
	\item The dimensionless coupling falls off faster than $k^2$. The anomalous scaling dominates the flow, and $Y_k \to 0$. The Ostrogradski ghost decouples and the theory is stable.
	\item The dimensionless coupling converges exactly quadratically. The dimensionful coupling $Y_k$ approaches a constant, which can be either zero or nonzero. The theory is stable only if this constant is zero.
\end{enumerate}
We will now discuss the IR behavior of the several classes of trajectories. Most of the physically interesting trajectories fall into the classes Type Ia, Type IIa, or Type IIIa introduced in Fig.\ \ref{fig:kganomdim}. The only trajectories which are not captured by this classification are the trajectories connecting the NGFPs which will be discussed separately. Our investigation reveals that the phase diagrams shown in Figs.\ \ref{fig:phasediagram} and \ref{fig:phasediagram2} realize all of the three cases described above.

\paragraph{Trajectories ending at the GFP (Type IIa).}
We start our analysis by considering Type IIa trajectories for which the cosmological constant $\Lambda_k$ flows to zero for $k\to 0$. In this case the IR completion of the trajectory is provided by the GFP \eqref{GFP2}. 
The IR attractive hypersurface of the GFP is spanned by the two eigenvectors associated with the negative stability coefficients $\theta_2 = \theta_3 = - 2$. The explicit expression for these eigenvectors are $e_1 = \hat{y}$ and $e_2 =\frac{2+N_s}{16\pi} \hat{\lambda} + \hat{g}$, where $\hat{y}$, $\hat{\lambda}$ and $\hat{g}$ are the unit vectors along the $y$, $\lambda$ and $g$-axis, respectively. By linearizing the flow at the GFP 
one finds that along these scaling directions
\be
y_k
=
y_{k_0}	\left(	\tfrac{k^2}{k_0^2}	\right)
\quad	\Leftrightarrow	\quad
Y_k
=
Y_{k_0}
\text{.}
\ee
Hence, there is a single RG trajectory, specified by $Y_{k_0} = 0$, for which $Y_0 = 0$ and the mass of the Ostrogradski ghost becomes infinite. This is the trajectory that has no initial component in the $\hat{y}$-direction, i.e. the one that approaches the GFP along $e_2$. Integrating the beta functions numerically one finds that this trajectory belongs to the UV-critical hypersurface of \NGFP{-}.

\begin{figure}[t!]
	\begin{center}
		\includegraphics{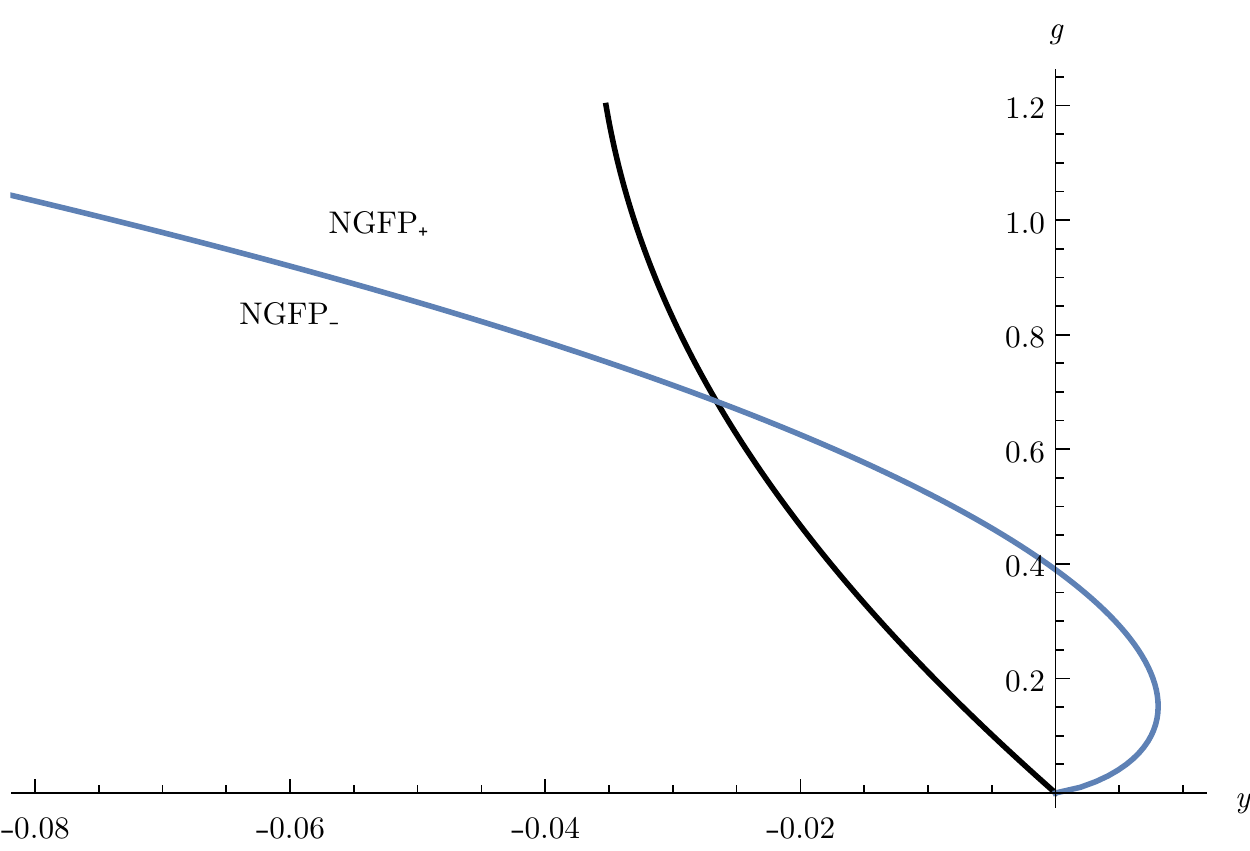}
	\end{center}
	\caption{Behavior of the RG trajectories passing through $y$-$g$-plane situated at $\lambda = -0.1$. Trajectories passing the plane above (below) the blue line emanate from \NGFP{+} (\NGFP{-}) while the high-energy behavior of trajectories building up the blue line is governed by the \NGFP{0}. Trajectories for which $\lim_{k \rightarrow 0} Y_k = 0$ are indicated by the black line.}
	\label{fig:unittrajphasediagcut}
\end{figure}

\paragraph{Trajectories of Type Ia and IIIa.}
Fig.\ \ref{fig:kganomdim} illustrates the existence of RG trajectories where $\lambda_k$ flows towards negative or positive infinity as $k \rightarrow 0$. The corresponding solutions are then classified as trajectories of Types Ia and IIIa, respectively. In order to determine the IR behavior of these trajectories, we numerically integrate the beta functions. Trajectories of Type IIIa terminate at $\eta_N^{\rm sing}$ at a finite value of $k$ and can not be completed to $k=0$ in the present approximation. Therefore, we limit our analysis to trajectories of Type Ia which extend up to $k=0$. The IR values $Y_0 \equiv \lim_{k \rightarrow 0} Y_k$ arising within this class of solutions are conveniently illustrated by studying the behavior of RG trajectories piercing the $y$-$g$-plane located at $\lambda = -0.1$ since the flow is essentially perpendicular to this plane. The resulting structure is illustrated in Fig.\ \ref{fig:unittrajphasediagcut}.
The plot shows that Type Ia trajectories can emanate from all three NGFPs: trajectories coming from \NGFP{0} pass the plane at the blue line while trajectories above (below) this line lie in the UV-critical surface of \NGFP{+} (\NGFP{-}). Trajectories where $Y_0 = 0$ span the black line in this diagram. Thus there is a 1-dimensional surface of solutions where the renormalized squared mass of the Ostrogradski ghost, \eqref{renghostmass}, is infinite such that the resulting degree of freedom does not propagate. Imposing the physical requirement that the renormalized scalar propagator does not give rise to an Ostrogradski ghost may then be used to fix one of the free parameters of the theory from stability considerations.

\paragraph{Trajectories flowing to \NGFP{0}.}
The final option for taking an IR limit consists in approaching \NGFP{0} along its IR-attractive eigendirection. From Fig.\ \ref{fig:phasediagram} one sees that there is a one-parameter family of trajectories emanating from either \NGFP{\pm} which end at \NGFP{0} as $k\rightarrow 0$. Linearizing the RG flow at the \NGFP{0} and using the stability coefficient along the IR attractive eigendirection listed in Table \ref{tab:ns1fpdata} yields the RG evolution of $y_k$ for these trajectories:
\be
\lim_{k \rightarrow 0} y_k
=
\left(	\tfrac{k}{k_0}	\right)^{0.529}	y_{k_0}
\qquad	\implies	\qquad
Y_k
=
\left(	\tfrac{k_0}{k}	\right)^{1.471}	Y_{k_0}
\text{.}
\ee
Since the scaling of the dimensionless $y$ is significantly smaller than $k^2$, the dimensionful $Y$ diverges as $k\to 0$ for all initial values $y \neq 0$. As a consequence the IR value of the ghost mass vanishes and the two terms describing the propagation of the scalar field in eq.\ \eqref{pfd} mutually cancel. Loosely speaking, the physical degree of freedom is eaten by the ghost so that the scalar does not propagate anymore. Verifying the robustness of this cancellation-mechanism  requires the inclusion of further powers $p^6, p^8, \ldots$ in the scalar propagator. This analysis is beyond the scope of the present work, however, and will be addressed in a forthcoming publication \cite{us:toappear}. 

\section{Conclusions and outlook}
\label{conclusion}
In this work, we use the effective average action $\Gamma_k$ to study the renormalization group flow of gravity coupled to scalar matter. Our ansatz for $\Gamma_k$ is given by the Einstein-Hilbert action coupled to an arbitrary number of minimally coupled scalar fields. The novel feature of the setup is the inclusion of a higher-derivative term in the scalar propagator. At the classical level these types of actions suffer from the so-called Ostrogradski instability reviewed in Sect.\ \ref{sect.2}: the appearance of degrees of freedom with a wrong-sign kinetic term, so-called Ostrogradski ghosts, renders the theory either unstable or non-unitary. At the same time it is clear that a generic RG flow will generate such potentially dangerous higher-derivative terms dynamically. This work initiates the systematic study of these types of terms in the RG framework with the goal of assessing their hazard potential for asymptotically safe theories. 

The quantity that actually encodes the relevant information on the spectrum of the theory is the \emph{renormalized propagator}. Exploiting that the effective average action obeys $\lim_{k \rightarrow 0} \Gamma_k = \Gamma$ with $\Gamma$ being the  standard effective action, this quantity can be accessed in the IR-limit of the flow. Within the present approximation the stability properties of the theory are captured by the IR-value of the (squared) Ostrogradski ghost mass $\mu^2 = Y_0^{-1}$. The ghost decouples from the spectrum if $Y_0 = 0$, so that the setting may give rise to stable (or equivalently unitary) theories even though the generic actions include higher-derivative kinetic terms.

The detailed study of the RG flow then established the following picture. In absence of the higher-derivative term the setting gives rise to a unique non-Gaussian fixed point (NGFP) suitable for rendering the gravity-matter system asymptotically safe. Upon including the scale-dependent Ostrogradski ghost mass, this NGFP splits up into three NGFPs which are labeled by the sign of $Y_*$. Notably there is one fixed point solution \NGFP{0} for which $Y_* = 0$ for all values of $k$. 

When projected to the $\lambda$-$y$-$g$-plane (see Figs. \ref{fig:phasediagram} and \ref{fig:phasediagram2}) the system of NGFPs essentially possesses a UV-critical hypersurface with three relevant directions. Within this space we have identified a two-dimensional subspace of RG trajectories that have a ghost-free IR limit. Phrased differently, the Ostrogradski ghost mass corresponds to a relevant direction of the NGFPs coming with a new free parameter. This freedom can be fixed by the requirement that the theory should contain only physical degrees of freedom in the IR. In this way the construction elegantly circumvents the potential danger of Ostrogradski instabilities by introducing a new free parameter and a mechanism to fix its value simultaneously. The analysis in Sect.\ \ref{sect.4} shows that the set of complete, unitary RG trajectories obtained from the full $\lambda$-$g$-$y$-system \eqref{beta1} is in one-to-one correspondence with the one found in the reduced system excluding the higher-derivative coupling.

As a byproduct, our analysis also provided new insights on potential bounds on the number of scalar fields compatible with the asymptotic safety mechanism. Throughout the calculation, we used a coarse graining operator of Type I (see \cite{Codello:2008vh} for an extended discussion), and extracted the running of $\eta_N$ from the background Newton's constant. The resulting analysis indicates that there are NGFPs suitable for realizing asymptotic safety for all values $N_s$. The characteristic fixed point properties shown in Fig.\ \ref{fig:kgfplocation} are strikingly similar to the ones found for foliated gravity-matter systems \cite{Biemans:2017zca}. Notably, our results also agree with the ones reported in \cite{Dona:2013qba}, where an upper bound $N_s \lesssim 17$ has been obtained. The crucial difference between the two settings lies in the choice of coarse-graining operator in the gravitational sector: our analysis uses a Type I coarse-graining operator while \cite{Dona:2013qba} resorts to a coarse-graining operator of Type II. If the analysis of Sect.\ \ref{sect.4a} is repeated for a coarse-graining operator of Type II, which effectively replaces eq.\ \eqref{B1TypeI} by \eqref{B1TypeII},  the reduced system \eqref{beta1} gives rise to the same upper bound on the number of scalar fields $N_s \lesssim 17$. From Fig.\ \ref{fig:sing} one then expects that the singular line $\eta_N^{\rm sing}$ plays a decisive role in stabilizing the NGFP for large values $N_s$.

Our analysis demonstrates that the existence of unitary RG trajectories is a non-trivial feature. A priori, a kinetic function of polynomial type is bound to have multiple roots, yielding a ghost in the particle spectrum. Further investigations suggesting themselves include studying a) polynomial truncations including further powers of the momentum,  $p^6, p^8,\ldots$, or b) truncations of non-polynomial type. In the first case, a higher-order truncation allows to investigate whether RG properties in lower orders are stable. In the second case, non-polynomial kinetic functions open up the possibility to have analytic kinetic functions without multiple roots, giving a ghost-free spectrum. An example is a propagator of the type $\mathrm{e}^{-\Delta} (\Delta + m^2)^{-1}$ studied, e.g. in the context of non-local gravity models \cite{Modesto:2011kw,Modesto:2014lga,Giaccari:2015vfh,Modesto:2017sdr}.\footnote{For earlier works on non-local gravity also see \cite{Biswas:2005qr,Biswas:2006bs,Biswas:2011ar,Talaganis:2014ida}. We thank A.\ Mazumdar for pointing out these references.} The unitarity conditions on such kinetic functions are studied in a separate paper \cite{mathphyscollaboration} and we hope to come back to the other points in the near future as well.

\medskip
\paragraph*{Acknowledgements.}
We thank R.\ Alkofer, F.\ Arici, J.\ Pawlowski, M.\ Reuter, W.\ van Suijlekom, and C.~Wetterich for helpful discussions. The research of D.~B., C.~R. and F.~S.
is supported by the Netherlands Organisation for Scientific
Research (NWO) within the Foundation for Fundamental Research on Matter (FOM) grant 13VP12.

\begin{appendix}
\section{Expanding trace arguments including step functions}
\label{App.A}
In this appendix, we collect the technical details underlying the derivation of the beta functions in the scalar sector. In this case, it is most convenient to choose a flat background spacetime were $\gb_{\mu\nu} = \delta_{\mu\nu}$. This allows to use momentum space techniques to evaluate the diagrams shown in Fig.\ \ref{fig:feynman}.

\subsection{Explicit form of vertex functions and propagators}
We start by deriving the relevant propagators and interaction vertices from the ansatz for the effective average action \eqref{ans1}. The result is conveniently expressed in terms of the variations $\Gamma_k^{(k,l;m)}$ where the number of derivatives with respect to the metric fluctuations and scalar fluctuations are denoted by $k$ and $l$, respectively. The number $m$ denotes the number of remaining background scalar fields. Moreover, we use the index $w$ to specify whether the building block is associated with the graviton ($w = hh$) or scalar fluctuations ($w = \phi\phi$). 

By expanding the gravitational sector up to second order in $h_{\mu\nu}$ one finds that the (inverse) gravitational propagator is given by
\be\label{gravprop}
\left[\Gamma_k^{(2,0;0)}\right]^{\mu\nu,\alpha\beta} = \frac{1}{32 \pi G_k} \left(p^2 - 2 \Lambda_k \right) \left[ (\unit - P_h) - \tfrac{d-2}{2} P_h \right]^{\mu\nu,\alpha\beta}
\ee
where $\unit_{\mu\nu}{}^{\alpha\beta} \equiv \tfrac{1}{2} \left( \delta_{\mu}^\alpha \delta_{\nu}^\beta + \delta_{\mu}^\beta \delta_{\nu}^\alpha \right)$ is the unit on the space of symmetric tensors and $[P_h]_{\mu\nu}{}^{\alpha\beta} \equiv d^{-1} \delta_{\mu\nu} \delta^{\alpha\beta}$ the projector on the trace mode. The (inverse) scalar propagator is obtained from \eqref{scalaraction} and reads
\be\label{scalarprop}
\Gamma_k^{(0,2;0)} = Z_k \left(p^2 + Y_k \, p^4 \right) \, . 
\ee
For later convenience, we introduce the following short-hand notations for the scale-dependent coefficients $\alpha_n^w$ multiplying the $p^{2n}$ terms in the (scalar part) of eqs.\ \eqref{gravprop} and \eqref{scalarprop},
\be
\begin{array}{lll}
	\alpha_0^{hh}=-\tfrac{\Lambda_k}{16\pi G_k}\,,\qquad &
	\alpha_1^{hh}=\tfrac{1}{32\pi G_k} \,,\qquad &
	\alpha_2^{hh}=0 \,, \\
	\alpha_0^{\phi\phi}=0 \,,\qquad &
	\alpha_1^{\phi\phi}=Z_k \,,\qquad &
	\alpha_2^{\phi\phi}=Z_k Y_K \, ,
\end{array}
\ee
and all coefficients $\alpha_n^w$ with $n \ge 3$ vanishing.  

In addition to the propagators, one also needs the (momentum-dependent) three- and four-point vertices containing one and two derivatives with respect to the background scalar field. Denoting the momenta associated with the graviton fluctuations, scalar fluctuations, and background scalar field by $\tilde{p}$, $p$,  and $q$, respectively the 3-point vertex obtained from \eqref{scalaraction} is
\be\label{3ptvtx}
\left[	\Gamma^{(1,1,1)}(\tilde{p},p,q)	\right]^{\mu\nu}
= Z_k \left( p^{(\mu} q^{\nu)} - \tfrac{1}{2} \delta^{\mu\nu} (p \cdot q) \right) .
\ee
Finally, the 4-point vertex is
\be\label{4ptvtx}
\left[	\Gamma^{(2,0,2)}	\right]^{\mu\nu,\rho\sigma}
= - \tfrac{1}{2}
Z_k	\left[
\left(
\tfrac{1}{4}	\delta^{\mu\nu}	\delta^{\rho\sigma}
-	\tfrac{1}{2}	\delta^{\mu\rho}	\delta^{\nu\sigma}
\right)	(q_1 \cdot q_2)
-	 \bar{g}^{\mu\nu}	q^\rho_1	q^\sigma_2
+	2	\bar{g}^{\mu\rho}	q^\sigma_1	q^\nu_2
\right]
\text{.}
\ee
All vertices are understood to contain the appropriate symmetrizations in the external indices and are subject to momentum conservation. Moreover, we set $Y_k = 0$ in order to keep the expressions for the vertices at a readable length. The contributions proportional to $Y_k$ are easily generated by a computer algebra program. Their precise form is irrelevant for the discussion of the general structures below.

Applying the implicit regulator prescription $p^2 \mapsto P_k = p^2 + R_k(p^2)$ to the propagators \eqref{gravprop} and \eqref{scalarprop} 

\begin{subequations}\label{regulatorsexp}
	\begin{align}\label{reghh}
	\left[\cR_k^{hh}\right]^{\mu\nu,\alpha\beta} = & \, \frac{1}{32 \pi G_k} \, R_k \, \left[ (\unit - P_h) - \tfrac{d-2}{2} P_h \right]^{\mu\nu,\alpha\beta} \\ \label{regpp}
	\cR_k^{\phi\phi} = & \, Z_k \left(1 + Y_k \left(2 \, p^2 + R_k \right) \right)  \, R_k .
	\end{align}
\end{subequations}
For the Litim-type cutoff \cite{Litim:2000ci,Litim:2001up} the dimensionful profile function $R_k$ is given by
\be\label{Litimreg}
R_k(p^2) = (k^2 - p^2) \Theta(k^2-p^2) \, . 
\ee
The key advantage of this regulator is that it allows for an analytic evaluation of the loop integrals shown in Fig.\ \ref{fig:feynman}. The distributional character of the regulator renders the expansion in the external momenta $q$ non-trivial, however. The next subsection discusses how this expansion can be implemented consistently, also taking into account the non-trivial boundary terms arising in the expansion procedure.

\subsection{Loop-integrations with a distributional regulator}
The loop integrals entailed by Fig.\ \ref{fig:feynman} contain a trace over spacetime indices and an integration over loop momenta.\footnote{We adopt the conventions that the absolute values of the loop momentum and external momentum  are denoted by $p$ and $q$ and $p \cdot q = pq \cos(\vartheta)$ defines their relative angle $\vartheta$. Moreover, the loop momentum is parameterized such that the external momentum enters into the propagator without regulator insertion only.} The spacetime indices are taken into account by stringing together the propagators and vertices contracting the corresponding index structures. This results in $q$-dependent scalar loop-integrals of the form
\be\label{eqn:masterIntegral}
I_{(m,n)}^{w_1w_2}
\equiv
\int \frac{d^d p}{(2\pi)^d} \, \frac{F_k(q,p,\cos (\vartheta))\,  \partial_t  \mathcal{R}^{w_2}_k(p^2) }{
	\left(\sum_{l=0}^2 \alpha_l^{w_1} (\vec{p}+\vec{q})^{2l} +\mathcal{R}^{w_1}_k((\vec{p}+\vec{q})^2)\right)^m
	\left(\sum_{l=0}^2 \alpha_l^{w_2} p^{2l} +\mathcal{R}^{w_2}_k(p^2)\right)^n
}\,.
\ee
Here $m$ and $n$ encode the number of propagators appearing in the diagram and, in a slight abuse of notation, the symbol $\mathcal{R}^{hh}_k(p^2)$ is used to refer to the scalar part of \eqref{reghh}. Diagrams containing 3-point vertices have $(m,n) = (1,2)$ while the tadpole diagram comes with $(m,n) = (0,2)$. The function $F_k(q,p,\cos (\vartheta))$ captures the momentum dependence of the vertices and is polynomial in $q$ and $p$. In particular it has a well-defined series expansion around $q=0$. Noting that the vertices \eqref{3ptvtx} and \eqref{4ptvtx} come with one and two powers of the external momentum, respectively, it is easy to verify that this expansion starts at order $q^2$.

For a general profile function $R_k$ the integrals eq.\ \eqref{eqn:masterIntegral} cannot be computed analytically. Moreover, the presence of the external momentum $q$ and the scale-dependent couplings make their numerical evaluation computationally very expensive. The profile function \eqref{Litimreg} allows to bypass this problem by restricting the $p$-integration to a compact domain and giving rise to cancellations in the propagators. The former property can be verified by noting that the logarithmic $k$-derivative of \eqref{regulatorsexp}, evaluated for a Litim profile, has the form
\be
\p_t \cR_k^w(p^2) = \bar{b}_k^w(p^2) \, \Theta(k^2 - p^2)
\ee
where
\be
\begin{split}
	\bar{b}_k^{hh}(p^2) = & \, \frac{1}{32 \pi G_k} \, \left( 2k^2 - \eta_N \left( k^2 - p^2 \right) \right)   \, ,  \\
	\bar{b}_k^{\phi\phi}(p^2) = & \, Z_k \left(2k^2 - \eta_s (k^2-p^2) + ( \p_t Y_k - \eta_s Y_k) (k^4-p^4) + 4 \, Y_k \, k^4 \right) \, . 
\end{split}
\ee

Inspecting \eqref{eqn:masterIntegral} for the case $m=0$ (tadpole diagram) reveals that  the step-functions appearing in the numerator and denominator have the same support. As a result the integrals simplify significantly
\be\label{simp1}
I_{(0,n)}^{w_1w_2}
\equiv \int d\Omega \int_{-1}^1 d\cos(\vartheta)
\int_0^k \frac{dp}{(2\pi)^d} \,p^{d-1} \,  \frac{F_k(q,p,\cos (\vartheta))\,  \bar{b}^w_k(p^2) }{
	\left(\sum_{l=0}^2 \alpha_l^w \, k^{2l} \right)^n
}\, .
\ee
Here $\int d\Omega$ denotes an angular integration and the spacetime indices on $F_k$ and $\bar{b}^w_k(p^2)$ are suppressed for readability. Owed to the simple structure of the denominator, which is independent of $p$ and $\vartheta$ the evaluation of these integrals is rather straightforward.

\begin{figure}[t]
	\begin{tikzpicture}[scale=0.8, every node/.style={scale=0.9}]
	\node[anchor=south] at (0,0){
		\tikz[scale=2]{
			\draw[draw=none,fill=gray!50] (-1,0) rectangle (1,1);
			\draw[] (1,0) node[anchor=north] {$1 $} -- node[anchor=north]{$\cos(\vartheta)$} (-1,0) node[anchor=north] {$-1 $} node[anchor=east] {0} --node[anchor=east] {$p^2$} (-1,1) node[anchor=east] {$k^2$} -- (1,1) --cycle;
			\node[anchor=south] at (0,1) {$q=0$};
		}
	};
	\node[anchor=south] at (6,0){
		\tikz[scale=2]{
			\draw[draw=none,fill=gray!50] (-0.25,1) to[out=-60, in=175] (1,0.25) -- (1,0) -- (-1,0) -- (-1,1) --cycle;
			\draw[] (1,0) node[anchor=north] {$1 $} -- node[anchor=north]{$\cos(\vartheta)$} (-1,0) node[anchor=north] {$-1 $} node[anchor=east] {0} --node[anchor=east] {$p^2$} (-1,1) node[anchor=east] {$k^2$} -- (1,1) --cycle;
			\node[anchor=south] at (0,1) {$q=\tfrac{1}{2}k$};
		}
	};
	\node[anchor=south] at (12,0){
		\tikz[scale=2]{
			\draw[] (1,0) node[anchor=north] {$1 $} -- node[anchor=north]{$\cos(\vartheta)$} (-1,0) node[anchor=north] {$-1 $} node[anchor=east] {0} --node[anchor=east] {$p^2$} (-1,1) node[anchor=east] {$k^2$} -- (1,1) --cycle;
			\node[anchor=south] at (0,1) {$q>2k$};
		}
	};
	\end{tikzpicture}
	\caption{\label{fig.stepfct} The value of $\Theta(k^2-(\vec{p}+\vec{q})^2)$ for three different values of $q$. In the gray regions the step function evaluates to $1$ while it vanishes in the white regions.}
\end{figure}
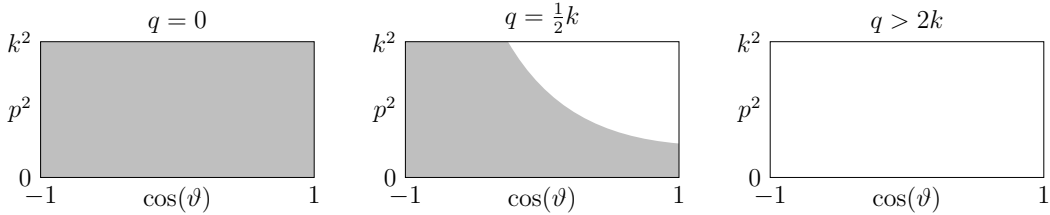

The case where $m \not = 0$ is non-trivial, however. Owed to the step-function in the numerator the full integration domain is reduced to a $d$-dimensional ball of radius $k$, i.e., $p\in[0,k]$ and $\cos(\vartheta)\in[-1,1]$. In this domain the second set of propagators again undergoes the simplification \eqref{simp1}. In the first set of propagators the regulator leads to terms proportional to $\Theta(k^2-(\vec{p}+\vec{q})^2 )$, however. As illustrated in Fig.\ \ref{fig.stepfct}, the value of the step function has a non-trivial dependence on the absolute value of $q$ and the angle $\vartheta$.  
Thus, unless $q=0$, there is always a part of the integration domain on which the denominator does not become trivial. As a result performing the integral becomes very involved. In order to complete the evaluation of the flow equation we then expand the integrands around $q=0$, \emph{taking the distributional character of the integrand into account}. This allows us to obtain analytic expressions for the resulting integrals. This is achieved as follows.

The first step uses the Heaviside function in the numerator of eq. \eqref{eqn:masterIntegral} to restrict the integration domain to $p\in [0,k]$. Following the derivation of \eqref{simp1} the factor $	\left(\sum_{l=0}^2 \alpha_l^{w_2} p^{2l} +\mathcal{R}^{w_2}_k(p^2)\right)^n$ becomes independent of $p$- and $q$. Together with the angular integration $\int {\rm d}\Omega$ it can then be absorbed into a prefactor 
\be
N_k^{n} \equiv 	\left(\sum_{l=0}^2 \alpha_l^w \, k^{2l} \right)^{-n} \, \int {\rm d}\Omega
\ee
so that eq.\ \eqref{eqn:masterIntegral} reduces to
\be
I_{(m,n)}^{w_1w_2} = 
N_k^{n}
\int_{-1}^{1}\!\!\! {\rm d}\cos(\vartheta)\!\int_{0}^{k} \!\!\!\!\frac{{\rm d} p}{(2\pi)^d} \, \frac{p^{d-1}\, \, F_k(q,p,\cos (\vartheta))\, \bar{b}_k^{w_2}(p^2) }{
	\left(\sum_{\ell=0}^{\infty}\alpha^{w_1}_{\ell}(\vec{p}+\vec{q})^{2\ell}+\mathcal{R}^{w_1}_k((\vec{p}+\vec{q})^2)\right)^m
} \, . 
\label{eqn:masterIntegral02}
\ee

In the next step we eliminate the step function from the denominator. For this purpose we insert the following partition of unity
\be\label{pou}
1=\Theta((\vec{p}+\vec{q})^2-k^2)+\Theta(k^2-(\vec{p}+\vec{q})^2) \, , 
\ee
defined in the weak sense. Furthermore we set $\Theta(0) = \tfrac{1}{2}$, so that this point is distributed evenly among the two terms. Inserting \eqref{pou} into \eqref{eqn:masterIntegral02} then gives
\be\label{eqn:masterIntegral03}
\begin{split}
	I_{(m,n)}^{w_1w_2} = & \, 
	N_k^{n}
	\int_{-1}^{1}\!\!\! {\rm d}\cos(\vartheta)\!\int_{0}^{k} \!\!\!\!\frac{{\rm d} p}{(2\pi)^d} \, p^{d-1}\, \, F_k(q,p,\cos (\vartheta))\, \bar{b}_k^{w_2}(p^2) \times \\
	&  \qquad \qquad \times
	\left\{
	\frac{\Theta((\vec{p}+\vec{q})^2-k^2)}{\left(\sum_{\ell=0}^{2}\alpha^{{w}_1}_{\ell}(\vec{p}+\vec{q})^{2\ell}\right)^n}
	+\frac{\Theta(k^2-(\vec{p}+\vec{q})^2)}{\left(\sum_{\ell=0}^{2}\alpha^{{w}_1}_{\ell}k^{2\ell}\right)^n		}
	\right\}
	\, .
\end{split}
\ee

The anomalous dimension of the scalar field and the beta function for $Y_k$ are encoded in the terms proportional to $q^2$ and $q^4$, respectively. For the present computation it thus suffices to expand \eqref{eqn:masterIntegral03} around $q=0$. The integral kernel itself is a distribution and thus its formal expansion yields distributional coefficients.
In a weak sense, which is suitable in the present context, the formal expansion coefficients can be constructed by using the integral representation of the Heaviside distribution
\begin{align}
\Theta(s)\equiv \lim_{\epsilon \to 0^+} \frac{1}{2\pi \imath} \int_{-\infty}^{\infty}{\rm d}t\, \left(t-\imath \epsilon\right)^{-1} e^{\imath t s}\,.
\end{align}
Substituting $s=(\pm(k^2-(\vec{p}+\vec{q})^2))$ and expanding the kernel in powers of $q$, we obtain, after taking the proper limit
\be\label{eqn:thetaExp001}
\begin{split}
	\Theta(\pm(k^2-(\vec{p}+\vec{q})^2)\,)
	\simeq & \, 
	\Theta(\pm(k^2-p^2))
	\mp	2 p \cos(\vartheta) \delta(\pm(k^2-p^2))\, q \\
	&
	+\big[	2 p^2 \cos^2(\vartheta) \delta^{\prime}(\pm(k^2-p^2)) \mp \delta(\pm(k^2-p^2)) \big]q^2 +\mathcal{O}(q^3) \, . 
\end{split}
\ee
Since the expansion of $F_k(q,p,\cos (\vartheta))$ starts at order $q^2$ it then suffices to terminate this expansion at order $q^2$. When inserting this representation into eq. \eqref{eqn:masterIntegral03} we encounter terms in which the delta-distribution has to be evaluated on the boundary of the integral domain.
Using $\Theta(0)\equiv \tfrac{1}{2}$, these can be evaluated by noting that
\be\label{wid1}
\int_{[0,a]}{\rm d}z\, G(z) \delta(a-z)\equiv \tfrac{1}{2}G(a)
\ee
which follows from
\be
\begin{split}
	G(a) 	=& \, G(0) +
	\int_{[0,a]}{\rm d}	z\, \partial_z G(z)\, \Theta(a-z) \\ 
	= & \, \tfrac{1}{2}G(a) + \int_{[0,a]}{\rm d}z\, G(z) \delta(a-z) \, . 
\end{split}
\ee
Finally, terms containing the $n$th derivative of the delta-function are evaluated using 
\begin{align}\label{wid2}
\int_{[0,1]} {\rm d}z\, G(z)\, \partial_s^n \delta(s)\big|_{s=\pm(1-z)} 
=& \, 
\tfrac{1}{2} (\pm 1)^{n} G^{(n)}(1) \, . 
\end{align}
The weak identities \eqref{wid1} and \eqref{wid2} are sufficient to derive the relevant trace contributions for the scalar beta functions.

\subsection{Master integrals}
\setlength{\belowdisplayskip}{3pt} \setlength{\belowdisplayshortskip}{0pt}
\setlength{\abovedisplayskip}{3pt} \setlength{\abovedisplayshortskip}{0pt}
We close the discussion by deriving a set of master integrals, which form the basis of our loop computations
\be\label{eqn:masterIntegral04}
\tilde{I}_{w,m}(q,\cos(\vartheta)) \equiv \int_{0}^{k} \!\!\!\! \frac{{\rm d} p}{(2\pi)^d} \, f(p) \, \left\{
\tfrac{\Theta((\vec{p}+\vec{q})^2-k^2)}{\left(\sum_{\ell=0}^{2}\alpha^{w}_{\ell}(\vec{p}+\vec{q})^{2\ell}\right)^m}
+\tfrac{\Theta(k^2-(\vec{p}+\vec{q})^2)}{\left(\sum_{\ell=0}^{2}\alpha^{w}_{\ell}k^{2\ell}\right)^m		}
\right\} \, .
\ee
Based on the relation \eqref{eqn:thetaExp001}, these integrals admit a series expansion in $q$,
\be
\tilde{I}_{w,m}(q,\cos(\vartheta)) \simeq \tilde{I}^{(0)}_{w,m} + \tilde{I}^{(1)}_{w,m} \, q + \tfrac{1}{2} \, \tilde{I}^{(2)}_{w,m} \, q^2 + \mathcal{O}(q^3) \, , 
\ee
where the series coefficients $\tilde{I}^{(n)}_{w,m}$ depend on $\cos(\vartheta)$. The first three coefficients in this expansion are found by substituting \eqref{eqn:thetaExp001} into \eqref{eqn:masterIntegral04} and evaluating the resulting integrals using the identities \eqref{wid1} and \eqref{wid2}
\begin{subequations}
	\begin{align}
	\tilde{I}^{(0)}_{w,m} = & \, \left(\sum_{\ell=0}^{2}\alpha^{w}_{\ell}k^{2\ell}\right)^{-m}\int_{0}^{k} \!\!\!\!\frac{{\rm d} p}{(2\pi)^d} \, f(p) \, , \\
	\tilde{I}^{(1)}_{w,m} = & \, 0 \, , \\
	\tilde{I}^{(2)}_{w,m} = & - m k^3 \cos^2(\vartheta)\left(\alpha^{w}_{1}+2\alpha^{w}_{2}k^2\right) \left(\sum_{l=0}^{2}\alpha^{w}_{l}k^{2l}\right)^{-(m+1)}  \frac{f(k)}{(2\pi)^{d}}	\,	.
	\end{align}
\end{subequations}
This result completes the discussion on carrying out the momentum integrals entailed by Fig.\ \ref{fig:feynman}. Note that the surface terms do not enter into the computation of the scalar anomalous dimension. They contribute to higher-order kinetic terms in the propagator only.
\end{appendix}



\begin{thebibliography}{99}

\bibitem{Niedermaier:2006wt}
M.~Niedermaier and M.~Reuter,
Living Rev.\ Rel.\  {\bf 9} (2006) 5.

\bibitem{Codello:2008vh}
A.~Codello, R.~Percacci and C.~Rahmede,
Annals Phys.\  {\bf 324} (2009) 414,
arXiv:0805.2909.

\bibitem{Litim:2011cp}
D.~F.~Litim,
Phil.\ Trans.\ Roy.\ Soc.\ Lond.\ A {\bf 369} (2011) 2759,
arXiv:1102.4624.



\bibitem{Reuter:2012id}
M.~Reuter and F.~Saueressig,
New J.\ Phys.\  {\bf 14} (2012) 055022,
arXiv:1202.2274.

\bibitem{Reuter:2012xf}
M.~Reuter and F.~Saueressig,
Lect.\ Notes Phys.\  {\bf 863} (2013) 185,
arXiv:1205.5431.

\bibitem{Nagy:2012ef}
S.~Nagy,
Annals Phys.\  {\bf 350} (2014) 310,
arXiv:1211.4151.


\bibitem{Roberto:book}
R.~Percacci,
\emph{An introduction to covariant quantum gravity and asymptotic safety},
World Scientific, Singapore (2017).

\bibitem{Bonanno:2017pkg}
A.~Bonanno and F.~Saueressig,
Comptes Rendus Physique {\bf 18} 254,
arXiv:1702.04137.

\bibitem{Weinberg:1980gg}
S.~Weinberg
in \textit{General Relativity, an Einstein Centenary Survey},
S.W.~Hawking and W.~Israel (Eds.),
Cambridge University Press, 1979; \\
S.~Weinberg,
hep-th/9702027.

\bibitem{Weinproc1}
S.~Weinberg, arXiv:0903.0568; PoS C {D09} (2009) 001, arXiv:0908.1964.

\bibitem{Reuter:1996cp}
M.~Reuter,
Phys.\ Rev.\ D {\bf 57} (1998) 971,
hep-th/9605030.

\bibitem{Gies:2016con}
H.~Gies, B.~Knorr, S.~Lippoldt and F.~Saueressig,
Phys.\ Rev.\ Lett.\  {\bf 116} (2016) 211302,
arXiv:1601.01800.

\bibitem{Reuter:2001ag}
M.~Reuter and F.~Saueressig,
Phys.\ Rev.\ D {\bf 65} (2002) 065016,
hep-th/0110054.



\bibitem{Eichhorn:2012va}
A.~Eichhorn,
Phys.\ Rev.\ D {\bf 86} (2012) 105021,
arXiv:1204.0965.



\bibitem{Dona:2013qba}
P.~Donà, A.~Eichhorn and R.~Percacci,
Phys.\ Rev.\ D {\bf 89} (2014) 084035,
arXiv:1311.2898.

\bibitem{Dona:2014pla}
P.~Donà, A.~Eichhorn and R.~Percacci,
Can.\ J.\ Phys.\  {\bf 93} (2015) 988,
arXiv:1410.4411.

\bibitem{Meibohm:2015twa}
J.~Meibohm, J.~M.~Pawlowski and M.~Reichert,
Phys.\ Rev.\ D {\bf 93} (2016) 084035,
arXiv:1510.07018.


\bibitem{Dona:2015tnf}
P.~Donà, A.~Eichhorn, P.~Labus and R.~Percacci,
Phys.\ Rev.\ D {\bf 93} (2016) 044049
Erratum: [Phys.\ Rev.\ D {\bf 93} (2016) 129904],
arXiv:1512.01589.





\bibitem{Biemans:2017zca}
J.~Biemans, A.~Platania and F.~Saueressig,
JHEP {\bf 05} (2017) 093,
arXiv:1702.06539.


\bibitem{Percacci:2002ie}
R.~Percacci and D.~Perini,
Phys.\ Rev.\ D {\bf 67} (2003) 081503,
hep-th/0207033.

\bibitem{Percacci:2003jz}
R.~Percacci and D.~Perini,
Phys.\ Rev.\ D {\bf 68} (2003) 044018,
hep-th/0304222.


\bibitem{Narain:2009fy}
G.~Narain and R.~Percacci,
Class.\ Quant.\ Grav.\  {\bf 27} (2010) 075001,
arXiv:0911.0386.

\bibitem{Narain:2009gb}
G.~Narain and C.~Rahmede,
Class.\ Quant.\ Grav.\  {\bf 27} (2010) 075002,
arXiv:0911.0394.

\bibitem{Benedetti:2013nya}
D.~Benedetti and F.~Guarnieri,
New J.\ Phys.\  {\bf 16} (2014) 053051,
arXiv:1311.1081.

\bibitem{Henz:2013oxa}
T.~Henz, J.~M.~Pawlowski, A.~Rodigast and C.~Wetterich,
Phys.\ Lett.\ B {\bf 727} (2013) 298,
arXiv:1304.7743.

\bibitem{Percacci:2015wwa}
R.~Percacci and G.~P.~Vacca,
Eur.\ Phys.\ J.\ C {\bf 75} (2015) 188,
arXiv:1501.00888.

\bibitem{Labus:2015ska}
P.~Labus, R.~Percacci and G.~P.~Vacca,
Phys.\ Lett.\ B {\bf 753} (2016) 274,
arXiv:1505.05393.

\bibitem{Borchardt:2016pif}
J.~Borchardt and B.~Knorr,
Phys.\ Rev.\ D {\bf 94} (2016) 025027,
arXiv:1603.06726.

\bibitem{Henz:2016aoh}
T.~Henz, J.~M.~Pawlowski and C.~Wetterich,
Phys.\ Lett.\ B {\bf 769} (2017) 105,
arXiv:1605.01858.


\bibitem{Bridle:2013sra}
I.~H.~Bridle, J.~A.~Dietz and T.~R.~Morris,
JHEP {\bf 1403} (2014) 093,
arXiv:1312.2846.


\bibitem{Benedetti:2014gja}
D.~Benedetti,
J.\ Stat.\ Mech.\  {\bf 1501} (2015) P01002,
arXiv:1403.6712.

\bibitem{Shapiro:2015ova}
I.~L.~Shapiro, P.~Morais Teixeira and A.~Wipf,
Eur.\ Phys.\ J.\ C {\bf 75} (2015) 262,
arXiv:1503.00874.



\bibitem{Guilleux:2015pma}
M.~Guilleux and J.~Serreau,
Phys.\ Rev.\ D {\bf 92} (2015) 084010,
arXiv:1506.06183; Phys.\ Rev.\ D {\bf 95} (2017) 045003, arXiv:1611.08106.



\bibitem{Robinson:2005fj}
S.~P.~Robinson and F.~Wilczek,
Phys.\ Rev.\ Lett.\  {\bf 96} (2006) 231601,
hep-th/0509050.


\bibitem{Pietrykowski:2006xy}
A.~R.~Pietrykowski,
Phys.\ Rev.\ Lett.\  {\bf 98} (2007) 061801,
hep-th/0606208.

\bibitem{Ebert:2007gf}
D.~Ebert, J.~Plefka and A.~Rodigast,
Phys.\ Lett.\ B {\bf 660} (2008) 579,
arXiv:0710.1002.

\bibitem{Toms:2007sk}
D.~J.~Toms,
Phys.\ Rev.\ D {\bf 76} (2007) 045015, 
arXiv:0708.2990; Phys.\ Rev.\ Lett.\  {\bf 101} (2008) 131301, arXiv:0809.3897; Phys.\ Rev.\ D {\bf 80} (2009) 064040, arXiv:0908.3100; Nature {\bf 468} (2010) 56, arXiv:1010.0793; Phys.\ Rev.\ D {\bf 84} (2011) 084016.







\bibitem{Daum:2009dn}
J.~E.~Daum, U.~Harst and M.~Reuter,
JHEP {\bf 01} (2010) 084,
arXiv:0910.4938.

\bibitem{Harst:2011zx}
U.~Harst and M.~Reuter,
JHEP {\bf 05} (2011) 119,
arXiv:1101.6007.

\bibitem{Folkerts:2011jz}
S.~Folkerts, D.~F.~Litim and J.~M.~Pawlowski,
Phys.\ Lett.\ B {\bf 709} (2012) 234,
arXiv:1101.5552.


\bibitem{Christiansen:2017gtg}
N.~Christiansen and A.~Eichhorn,
Phys.\ Lett.\ B {\bf 770} (2017) 154,
arXiv:1702.07724.

\bibitem{Eichhorn:2017lry}
A.~Eichhorn and F.~Versteegen,
arXiv:1709.07252 [hep-th].



\bibitem{Dona:2012am}
P.~Dona and R.~Percacci,
Phys.\ Rev.\ D {\bf 87} (2013) 045002,
arXiv:1209.3649.

\bibitem{Gies:2013noa}
H.~Gies and S.~Lippoldt,
Phys.\ Rev.\ D {\bf 89} (2014) 064040,
arXiv:1310.2509.


\bibitem{Eichhorn:2011pc}
A.~Eichhorn and H.~Gies,
New J.\ Phys.\  {\bf 13} (2011) 125012,
arXiv:1104.5366.

\bibitem{Meibohm:2016mkp}
J.~Meibohm and J.~M.~Pawlowski,
Eur.\ Phys.\ J.\ C {\bf 76} (2016) 285,
arXiv:1601.04597.

\bibitem{Eichhorn:2016vvy}
A.~Eichhorn and S.~Lippoldt,
Phys.\ Lett.\ B {\bf 767} (2017) 142,
arXiv:1611.05878.


\bibitem{Shaposhnikov:2009pv}
M.~Shaposhnikov and C.~Wetterich,
Phys.\ Lett.\ B {\bf 683} (2010) 196,
arXiv:0912.0208.

\bibitem{Eichhorn:2014qka}
A.~Eichhorn and M.~M.~Scherer,
Phys.\ Rev.\ D {\bf 90} (2014) 025023,
arXiv:1404.5962.

\bibitem{Eichhorn:2015kea}
A.~Eichhorn, H.~Gies, J.~Jaeckel, T.~Plehn, M.~M.~Scherer and R.~Sondenheimer,
JHEP {\bf 04} (2015) 022,
arXiv:1501.02812.

\bibitem{Eichhorn:2017ylw}
A.~Eichhorn and A.~Held,
arXiv:1707.01107.


\bibitem{Zanusso:2009bs}
O.~Zanusso, L.~Zambelli, G.~P.~Vacca and R.~Percacci,
Phys.\ Lett.\ B {\bf 689} (2010) 90,
arXiv:0904.0938.

\bibitem{Oda:2015sma}
K.~y.~Oda and M.~Yamada,
Class.\ Quant.\ Grav.\  {\bf 33} (2016) 125011,
arXiv:1510.03734.

\bibitem{Eichhorn:2016esv}
A.~Eichhorn, A.~Held and J.~M.~Pawlowski,
Phys.\ Rev.\ D {\bf 94} (2016) 104027,
arXiv:1604.02041.

\bibitem{Hamada:2017rvn}
Y.~Hamada and M.~Yamada,
arXiv:1703.09033.

\bibitem{Christiansen:2017qca}
N.~Christiansen, A.~Eichhorn and A.~Held,
arXiv:1705.01858 [hep-th].

\bibitem{Gies:2017zwf}
H.~Gies, R.~Sondenheimer and M.~Warschinke,
arXiv:1707.04394.






\bibitem{Litim:2014uca}
D.~F.~Litim and F.~Sannino,
JHEP {\bf 1412} (2014) 178,
arXiv:1406.2337.

\bibitem{Esbensen:2015cjw}
J.~K.~Esbensen, T.~A.~Ryttov and F.~Sannino,
Phys.\ Rev.\ D {\bf 93} (2016) 045009,
arXiv:1512.04402.

\bibitem{Bond:2016dvk}
A.~D.~Bond and D.~F.~Litim,
Eur.\ Phys.\ J.\ C {\bf 77} (2017) 429,
arXiv:1608.00519.

\bibitem{Bond:2017wut}
A.~D.~Bond, G.~Hiller, K.~Kowalska and D.~F.~Litim,
JHEP {\bf 08} (2017) 004,
arXiv:1702.01727 [hep-ph].


\bibitem{Horndeski:1974wa}
G.~W.~Horndeski,
Int.\ J.\ Theor.\ Phys.\  {\bf 10} (1974) 363.

\bibitem{Langlois:2015cwa}
D.~Langlois and K.~Noui,
JCAP {\bf 02} (2016) 034,
arXiv:1510.06930.

\bibitem{Motohashi:2016ftl}
H.~Motohashi, K.~Noui, T.~Suyama, M.~Yamaguchi and D.~Langlois,
JCAP {\bf 07} (2016) 033,
arXiv:1603.09355.


\bibitem{Woodard:2015zca}
R.~P.~Woodard,
Scholarpedia {\bf 10} (2015) 32243,
arXiv:1506.02210.

\bibitem{Crisostomi:2017aim}
M.~Crisostomi, R.~Klein and D.~Roest,
JHEP {\bf 06} (2017) 124,
arXiv:1703.01623.



\bibitem{Ostrogradsky:1850}
M.~Ostrogradski, Mem. Ac. St. Petersbourg {\bf VI 4} (1850) 385.

\bibitem{Weinberg:1995mt}
S.~Weinberg,
\emph{The Quantum theory of fields. Vol. 1: Foundations}, Cambridge Univ. Press, Cambridge (UK), 1995.

\bibitem{mathphyscollaboration}
F.\ Arici, D.\ Becker, C.\ Ripken, F.\ Saueressig, and W.\ van Suijlekom in preparation.

\bibitem{Motohashi:2014opa}
H.~Motohashi and T.~Suyama,
Phys.\ Rev.\ D {\bf 91} (2015) 085009,
arXiv:1411.3721.




\bibitem{Barnaby:2007ve}
N.~Barnaby and N.~Kamran,
JHEP {\bf 02} (2008) 008,
arXiv:0709.3968.

\bibitem{Barnaby:2008tc}
N.~Barnaby and N.~Kamran,
JHEP {\bf 12} (2008) 022,
arXiv:0809.4513.





\bibitem{Modesto:2011kw}
L.~Modesto,
Phys.\ Rev.\ D {\bf 86} (2012) 044005,
arXiv:1107.2403.

\bibitem{Modesto:2014lga}
L.~Modesto and L.~Rachwal,
Nucl.\ Phys.\ B {\bf 889} (2014) 228,
arXiv:1407.8036.

\bibitem{Giaccari:2015vfh}
S.~Giaccari, L.~Modesto, L.~Rachwal and Y.~Zhu,
arXiv:1512.06206 [hep-th].

\bibitem{Modesto:2017sdr}
L.~Modesto and L.~Rachwal,
Int.\ J.\ Mod.\ Phys.\ D {\bf 26} (2017) 1730020.

\bibitem{Biswas:2005qr}
T.~Biswas, A.~Mazumdar and W.~Siegel,
JCAP {\bf 03} (2006) 009,
hep-th/0508194.

\bibitem{Biswas:2006bs}
T.~Biswas, R.~Brandenberger, A.~Mazumdar and W.~Siegel,
JCAP {\bf 0712} (2007) 011,
hep-th/0610274.

\bibitem{Biswas:2011ar}
T.~Biswas, E.~Gerwick, T.~Koivisto and A.~Mazumdar,
Phys.\ Rev.\ Lett.\  {\bf 108} (2012) 031101,
arXiv:1110.5249.

\bibitem{Talaganis:2014ida}
S.~Talaganis, T.~Biswas and A.~Mazumdar,
Class.\ Quant.\ Grav.\  {\bf 32} (2015) 215017,
arXiv:1412.3467.

\bibitem{Benedetti:2009rx}
D.~Benedetti, P.~F.~Machado and F.~Saueressig,
Mod.\ Phys.\ Lett.\ A {\bf 24} (2009) 2233,
arXiv:0901.2984.

\bibitem{Codello:2010mj}
A.~Codello,
Annals Phys.\  {\bf 325} (2010) 1727,
arXiv:1004.2171.



\bibitem{Wetterich:1992yh}
C.~Wetterich,
Phys.\ Lett.\ B {\bf 301} (1993) 90.

\bibitem{Morris:1993qb}
T.~R.~Morris,
Int.\ J.\ Mod.\ Phys.\ A {\bf 9} (1994) 2411
[hep-ph/9308265].

\bibitem{Reuter:1993kw}
M.~Reuter and C.~Wetterich,
Nucl.\ Phys.\ B {\bf 417} (1994) 181.



\bibitem{Ohta:2016npm}
N.~Ohta, R.~Percacci and A.~D.~Pereira,
JHEP {\bf 1606} (2016) 115,
arXiv:1605.00454.

\bibitem{Ohta:2016jvw}
N.~Ohta, R.~Percacci and A.~D.~Pereira,
arXiv:1610.07991 [hep-th].


\bibitem{Manrique:2011jc}
E.~Manrique, S.~Rechenberger and F.~Saueressig,
Phys.\ Rev.\ Lett.\  {\bf 106} (2011) 251302,
arXiv:1102.5012.

\bibitem{Biemans:2016rvp}
J.~Biemans, A.~Platania and F.~Saueressig,
Phys.\ Rev.\ D {\bf 95} (2017) 086013,
arXiv:1609.04813.

\bibitem{Houthoff:2017oam}
W.~B.~Houthoff, A.~Kurov and F.~Saueressig,
Eur.\ Phys.\ J.\ C {\bf 77} (2017) 491,
arXiv:1705.01848.

\bibitem{Wetterich:2017ixo}
C.~Wetterich, Phys.\ Lett.\ B {\bf 773} (2017) 6,
arXiv:1704.08040.



\bibitem{Litim:2000ci}
D.~F.~Litim,
Phys.\ Lett.\ B {\bf 486} (2000) 92,
hep-th/0005245.

\bibitem{Litim:2001up}
D.~F.~Litim,
Phys.\ Rev.\ D {\bf 64} (2001) 105007,
hep-th/0103195.


\bibitem{us:toappear}
D.\ Becker, C.\ Ripken and F.\ Saueressig, in preparation.


\end{thebibliography}
\end{document}